\documentclass[%
 reprint,
 amsmath,
 amssymb,
 aps,
 longbibliography,
]{revtex4-1}

\usepackage{graphicx}% Include figure files
\usepackage{dcolumn}% Align table columns on decimal point
\usepackage{bm}% bold math
\usepackage{hyperref}
\usepackage{braket}
\newcommand*{\affmark}[1][*]{\textsuperscript{#1}}

\begin{document}

%\preprint{APS/123-QED}
%DJR Edits. Nov 20, 2018.
\title{Transmission Lines and Meta-Materials based on Quantum Hall Plasmonics}

\author{S. Bosco\affmark[1,3]}
\email {bosco@physik.rwth-aachen.de}
\author{D. P. DiVincenzo\affmark[1,2,3]}
 \email{d.divincenzo@fz-juelich.de}
\author{D. J. Reilly\affmark[4,5]}
 \email{david.reilly@sydney.edu.au}

\affiliation{
\affmark[1]Institute for Quantum Information, RWTH Aachen University,                                
  D-52056
  Aachen,                              
  Germany
}

\affiliation{
  \affmark[2]Peter Gr\"{u}nberg Institute, Theoretical Nanoelectronics,
    Forschungszentrum J\"{u}lich,
  D-52425
  J\"{u}lich,
  Germany
}

\affiliation{
\affmark[3]J\"{u}lich-Aachen Research Alliance (JARA),
    Fundamentals of Future Information Technologies,
  D-52425
  J\"{u}lich,
  Germany
}

\affiliation{
\affmark[4] ARC Centre of Excellence for Engineered Quantum Systems, School of Physics, The University of Sydney, Sydney, NSW 2006, Australia.
 }

\affiliation{\affmark[5] Microsoft Quantum Sydney, The University of Sydney, Sydney, NSW 2006, Australia.
} 

\date{\today}

\begin{abstract}
The characteristic impedance of a microwave transmission line is typically constrained to a value $Z_0$ = 50 $ \Omega$, in-part because of the low impedance of free space and the limited range of permittivity and permeability realizable with conventional materials. Here we suggest the possibility of constructing high-impedance transmission lines by exploiting the plasmonic response of edge states associated with the quantum Hall effect in gated devices. We analyze various implementations of quantum Hall transmission lines based on distributed networks and lumped-element circuits, including a detailed account of parasitic capacitance and Coulomb drag effects, which can modify device performance. We additionally conceive of a meta-material structure comprising arrays of quantum Hall droplets and analyze its unusual properties. The realization of such structures holds promise for efficiently wiring-up quantum circuits on chip, as well as engineering strong coupling between semiconductor qubits and microwave photons.
\\

\end{abstract}

\pacs{Valid PACS appear here}% PACS, the Physics and Astronomy
                             % Classification Scheme.
\keywords{Quantum Hall effect, Edge magnetoplasmons, 2DEG, High impedance transmission line, Singlet-triplet qubits}%Use showkeys class option if keyword
                              %display desired
\maketitle

%\tableofcontents

\section{\label{sec:intro}Introduction}

Specifying the impedance of radio-frequency or microwave circuits greatly simplifies their analysis by making use of scattering matrices, rather than geometry-specific solutions to Maxwell's equations \cite{Rad_Lab}. Motivated by practical aspects, the characteristic impedance of transmission lines is today largely standardized to a value of $Z_0 = $ 50 or 75 $\Omega$, enabling the seamless integration of electrical components. Forgoing practicality however, much more fundamental considerations suggest that $Z_0$ cannot be set too far from this value. The fine structure constant, for instance, establishes the impedance of free-space at ($Z$ = 377 $\Omega$), with dielectrics or magnetic materials then modifying the characteristic impedance by a limited amount, commensurate with their relative permittivity or permeability. Going beyond these constraints, the large inductance of an array of superconducting Josephson junctions has recently been exploited to yield microwave resonators and transformers with impedances in the few kilo-Ohms \cite{doi:10.1063/1.4832074, Wallraff} and devices based on surface wave propagation in carbon nanotubes have been proposed \cite{1406008}. \\

For circuits operating in the mesoscale or quantum domain, the impedance of a conductor supporting a single ballistic mode is given by the quantum of resistance, defined by the von Klitzing constant, $R_K \sim 25.8 \mathrm{k\Omega}$ \cite{PhysRevLett.45.494}, which is far from typical values of $Z_0$ used in microwave engineering and above what has been realized with recent superconducting implementations \cite{doi:10.1063/1.4832074, Wallraff}. If it were possible to make use of ballistic conductors to establish high impedance transmission lines, they would provide a means of efficiently wiring-up quantum circuits on chip without the use of bulky, narrow-band impedance transformers, which limit, for example, the performance of qubit readout detectors \cite{Reilly:2007}. In fact, a high impedance also leads to a high voltage per photon, and consequently can enable enhanced electrostatic coupling between distributed resonator structures and qubits. This enhancement is particularly appealing for semiconductor-based quantum computing, where the qubits generally have an inconveniently small charge dipole, making it hard to achieve the strong coupling regime, where the photon-qubit coupling strength is higher than the losses in the resonator or the qubit.\\

In this paper, we propose and analyze low-loss, high-impedance microwave transmission lines and resonator structures realized using the plasmonic response of a system in the quantum Hall  (QH) regime, where transport is supported by only a few conducting channels. To avoid dissipation, the QH material cannot be ohmically contacted to the external electrodes, but instead it should be reactively coupled  \cite{Viola-DiVincenzo,Wick,Wick-patent,Girvin}. Here, we consider only a capacitive coupling between the electrodes and the quantum Hall material where transport is associated with edge magnetoplasmons (EMPs) - charge density excitations that travel with a velocity some 1000-times slower than the speed of light in vacuum \cite{Volkov,Glazman,JohnsonVignale,Glazman2,Circuit_EMP,QEMP}. \\

Realizing transmission line structures extends the tool-kit of useful quantum Hall devices available to address the challenges of quantum information processing \cite{Stace1,Landig,Benito-exp,Doherty1,Doherty2}. The chirality of these devices, for example, can be exploited to implement minaturized, scalable non-reciprocal devices such as gyrators and circulators \cite{Viola-DiVincenzo,Placke,Reilly,Reilly2} that are broadly used  for manipulation of qubits and back-action mitigation. Other passive implementations are also possible \cite{Stace2}. \\

In what follows we adopt a simple phenomenological model inspired by \cite{Viola-DiVincenzo}, and analyze the physics of quantum Hall based transmission lines. We discuss various possible implementations using gate electrodes, including compact devices that mimic branching transmission lines or `stubs'. These interferometric structures can be tuned to create on-chip impedance matching networks and compact delay-lines. Our model is extended to account for effects  associated with parasitic capacitance and interacting edge states that produce Coulomb drag. In addition, we show how by cascading patterns of quantum Hall droplets a new kind of meta-material can be realized with exotic bandstructure. These chains of cascaded droplets enable transmission lines of arbitrary length and shape. Of further interest,  for frequencies that correspond to the band-gap of the meta-material, transmission abruptly drops to zero, analogous to perfect Bragg reflection in a crystal. Such devices may have application in creating compact on-chip microwave filters with non-reciprocal properties. 

%A detailed microscopic justification of the phenomenological model used here is reported separately in Ref. \cite{HITLpt2}.

\section{Distributed transmission line\label{sec:distributed-tl}}

In general, the motion of the excess charge density $\rho$ localized at the edge of a dissipationless quantum Hall material and moving along the perimeter is described in the frequency domain by \cite{Viola-DiVincenzo,QEMP}:
\begin{equation}
\label{eq:motion-emp-local-velocity-app}
i\omega\rho(y,\omega )=  \partial_y\left(v(y) \rho(y,\omega)\right)+\sigma_{xy} \partial_y V_a(y,\omega),
\end{equation}
where  $V_a(y,\omega)$ is the time dependent drive applied, $\sigma_{xy}$ is the off-diagonal conductivity of the QH material and $v$ is the propagation velocity of the EMP as function of position $y$, possibly accounting for different screening of Coulomb interactions due to the presence of the electrodes.
A detailed discussion of the validity of this model can be found in \cite{QEMP,HITLpt2}. \\

The total current flowing into the $i$th electrode is obtained by integrating the displacement current density over its area. 
In the model presented here, we neglect fringing fields and, because the EMP charge density is assumed to be localized in an infinitesimally narrow stripe along the edge, the integral over the area of the electrode simplifies into the one dimensional integral \cite{QEMP,HITLpt2}:

\begin{equation}
\label{eq:current-ith-electrode}
I_i(\omega)=-i\omega\int_{l_i}^{r_i}dy\rho(y,\omega ),
\end{equation}
where $l_i$ and $r_i$ are respectively the left and right edges of the $i$th electrode. \\

\begin{figure}
\includegraphics[width=0.45\textwidth]{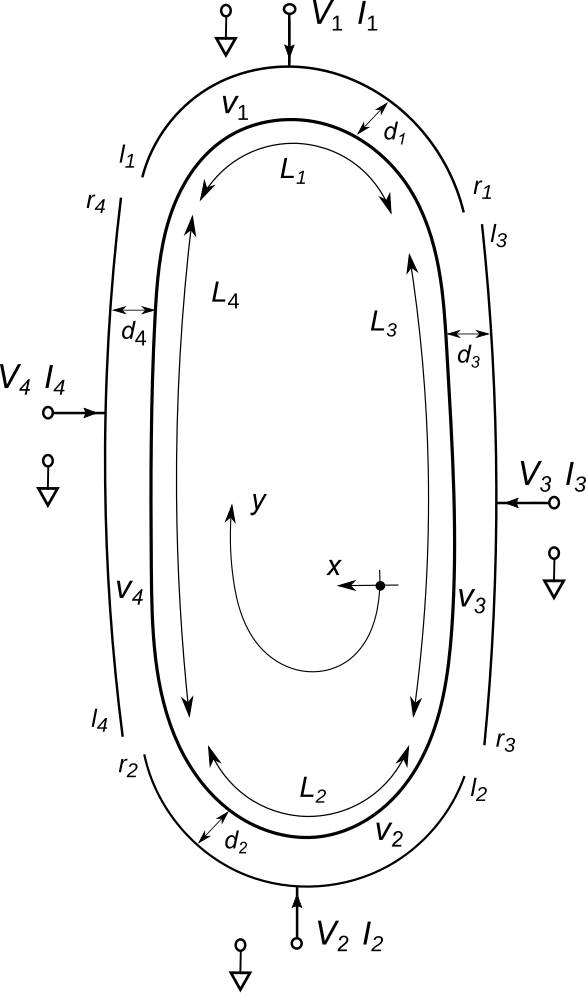}
\caption{Sketch of a quantum Hall transmission line. Four ideal electrodes of length $L_i$ are placed at a distance $d_i$ from the edge of a 2 dimensional QH droplet; the capacitively coupling between the material and the $i$th lead is quantified by the velocities $v_i$. 
Each of the electrode can in principle be driven by a voltage $V_i$ applied with respect to ground or by a current $I_i$.
The position in the material is parametrized by the coordinates $(x,y)$, that are respectively normal and tangential to the perimeter. $l_i$ and $r_i$ indicate the left and right edges of the $i$th electrode, respectively. We refer to the pair of electrodes 1,2 (3,4) as driving (screening) electrodes.  } 
\label{fig:transmission-line}
\end{figure}

In this section, we focus on the device shown in Fig.  \ref{fig:transmission-line}.
A QH droplet is capacitively coupled to four ideal metal electrodes of length $L_i$ that are placed at a distance $d_i$ from the edge. In principle, each of these electrode can be driven independently by a voltage $V_i$ applied with respect to ground, however we assume that only terminals 1 and 2 can be externally driven, while the others are not connected directly to a source, but they can either be grounded or left floating. Consequently, we will refer to the pairs 1,2 and 3,4 respectively as driving and screening electrodes.
Also, we neglect for now the parasitic capacitive coupling between the electrodes, that can be straightforwardly included a posteriori and whose effects are discussed in Sec. \ref{sec:parasitics}.

% We have a quantum Hall droplet capacitively coupled to two electrodes of length $L_{1,2}$, separated by distance $L_{3,4}$. To avoid logarithmic singularities in the EMP velocity, we assume the presence of two lateral electrodes in these regions. For simplicity, we consider these two electrodes to be placed symmetrically and we neglect for the moment the parasitic capacitive coupling between them, that can be straightforwardly included a posteriori. 
% These screening electrodes can be either grounded or left floating; to capture both cases in the analysis, we now assume that they can be independently driven by two independent voltage sources $V_{3,4}$. 

Note that in the setup chosen, all parts of the perimeter of the droplet are coupled to some external electrode. 
This choice guarantees that the EMP velocity has no logarithmic singularity in the long-wavelength limit \cite{Volkov}, and, when $d_i/L_i\ll 1$, it allows one to use a simple piecewise decomposition for the velocity function $v(y)$, with constant velocity $v_i$ in the region coupled to the $i$th electrode \cite{HITLpt2}. The velocity $v_i$ can then be estimated by using
\begin{equation}
\label{eq:EMP-vel-app}
v_i=\frac{\sigma_{xy}}{4\pi\epsilon_S} \log\left(1+4\frac{d_i^2}{l^2}\right),
\end{equation}
with $\epsilon_S$ being the dielectric constant of the medium and $l$ being the characteristic length over which the conductivity profile changes from zero to the bulk value; for example, in quantum Hall devices with atomically defined edges, $l$ corresponds approximately to the magnetic length $\sqrt{\hbar/(eB)}$  \cite{HITLpt2}. 
We find that  $v_i$ is of the order $10^5\mathrm{m/s}$, in agreement with  recent experiments in various materials \cite{Reilly,Glattli,Reilly2}.

With this piecewise approximation and using Eqs. (\ref{eq:motion-emp-local-velocity-app}) and (\ref{eq:current-ith-electrode}), one can compute a $4\times4$ terminal-wise admittance matrix with elements
\begin{subequations}
\label{eq:admittance-4x4}
\begin{flalign}
\begin{split}
Y_{ii}=& -\frac{\sigma_{xy}}{2}
 \left(1+i \cot \left(\frac{\omega}{2} \sum _{k} \tau_k\right)\right) \times\\
  &\   \   \   \   \   \   \   \   \left(1-e^{i \omega \tau_i}\right) \left(1-e^{i \omega \sum_{k \neq i} \tau_k}\right)
 \end{split},\\
 \begin{split}
 Y_{ij}=&\frac{\sigma_{xy}}{2}
 \left(1+i \cot \left(\frac{\omega}{2} \sum _{k} \tau_k\right)\right)\times\\
  &\   \   \   \   \   \   \   \   \left(1-e^{i \omega \tau _i}\right) \left(1-e^{i \omega \tau _j}\right)e^{i\omega \tau_{ij}^{\circlearrowleft}}.
 \end{split} 
\end{flalign}
\end{subequations}
This matrix relates the currents $(I_1,I_2,I_3,I_4)$ to the corresponding driving voltages measured with respect to a common ground.
We introduced the timescales 
\begin{equation}
\tau_i\equiv \frac{L_i}{v_i},
\end{equation}
that characterize the time spent by an EMP for traveling through the $i$th electrode and $\tau_{ij}^{\circlearrowleft}$ that specifies the total time required for traveling in the anticlockwise direction from the left edge of electrode $i$ ($l_i$) to the right edge of electrode $j$ ($r_j$), see Fig. \ref{fig:transmission-line}.
Note that this is a valid terminal admittance matrix, satisfying the requirements $\sum_i Y_{ij}=\sum_j Y_{ij}=0$, deriving from Kirchhoff's laws.

\begin{figure}
\includegraphics[width=0.45\textwidth]{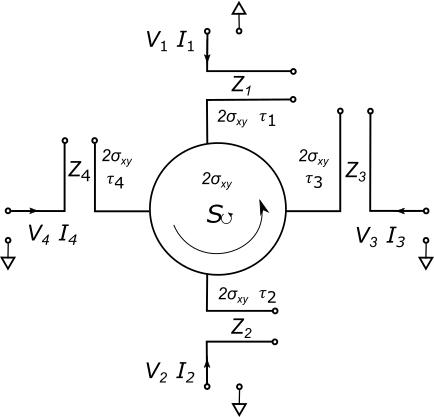}
\caption{Circuit equivalent of the QH droplet in Fig. \ref{fig:transmission-line}. The circuit is composed of an ideal anticlockwise circulator with characteristic impedance $1/(2\sigma_{xy})$ and scattering matrix $S_{\circlearrowleft}$ (Eq. (\ref{eq:S-circulator})) that connects in series four delay lines with frequency dependent impedance $Z_i$ (Eq. (\ref{eq:terminal-impedance-frequency})).
Each delay line is a transmission line with characteristic impedance $1/(2\sigma_{xy})$ and propagation time $\tau_i$ terminated by an open circuit, and it models the capacitive coupling between the QH material and the $i$th electrode.}
\label{fig:circuit-model}
\end{figure}

Now that we have a linear relation between applied voltages and currents, we can straightforwardly apply the preferred boundary conditions to the screening electrodes.
Before examining in detail the different situations, we can gain additional insights into the physics of these devices by noting that the admittance of the circuit model in Fig. \ref{fig:circuit-model} exactly reproduces Eq.   (\ref{eq:admittance-4x4}).
The finite time spent by an EMP for traveling through the $i$th electrode leads to the presence in this circuit model of delay lines with impedance
\begin{equation}
\label{eq:terminal-impedance-frequency}
Z_i(\omega)=-\frac{i}{2\sigma_{xy}}\cot\left(\frac{\omega\tau_i}{2}\right).
\end{equation}
This equation is the well-known input impedance of a TL terminated by an open circuit \cite{Pozar}.
Also, it is possible to show that if we add additional electrodes, this circuit model is straightforwardly generalized by adding the corresponding delay lines.

The chiral propagation of the EMP (in the anticlockwise direction when $v_i>0$) is captured by the presence in the model of an ideal circulator: a non-reciprocal device that cyclically routes the signal from one port to the next one in the direction fixed by the arrow.
With our unconventional ordering of the electrodes, the microwave scattering matrix of this component is
\begin{equation}
    \label{eq:S-circulator}
S_{\circlearrowleft} =
    \begin{pmatrix}
        0 & 0 & 1 & 0 \\
        0 & 0 & 0 & 1 \\
        0 & 1 & 0 & 0 \\
        1 & 0 & 0 & 0         
    \end{pmatrix},
\end{equation}
and it relates the outgoing voltage waves $V^{\text{out}}$ to the incoming ones $V^{\text{in}}$, defined for the $i$th terminal by
\begin{subequations}
\label{eq:voltage-waves-scattering}
\begin{flalign}
V_i&=V^{\text{in}}_i+V^{\text{out}}_i,\\
I_i&=\frac{1}{2\sigma_{xy}}(V^{\text{in}}_i-V^{\text{out}}_i).
\end{flalign} 
\end{subequations} 
We remark that while the $S$ matrix is generally used as a port matrix, here, for convenience, we use it as a terminal-wise quantity by measuring all the voltages $V_i$ with respect to a common ground. 

The characteristic impedance of the delay lines and of the circulator are proportional to $1/\sigma_{xy}$, and therefore, in the quantum Hall regime and  at low filling factor, they can be of the order of the quantum of resistance $R_K \sim 25.8 \mathrm{k\Omega}$.
Also, from Fig. \ref{fig:circuit-model}, it is evident that the full transmission of a signal from terminal 1 to terminal 2 is possible only at the frequencies for which the two delay lines $Z_{1,2}(\omega)$ act as a short:
\begin{equation}
\label{eq:cond_short}
Z_{1,2}(\omega_n)=0 \ \ \ \ \ \mathrm{i.e.} \ \ \ \ \frac{\omega_n \tau_{1,2}}{2\pi}=\frac{1}{2}+n ,
\end{equation}
while when they act as an open
\begin{equation}
\label{eq:cond_open}
Z_{1,2}(\omega_m)\rightarrow \infty \ \ \ \ \ \mathrm{i.e.} \ \ \ \ \frac{\omega_m \tau_{1,2}}{2\pi}=m ,
\end{equation}
the device would be perfectly reflecting (here $n,m\in \mathbb{N}_0$). Depending on how the screening gates are treated, the response can change drastically; we now examine in more detail different situations and we verify when the device can mimic a conventional TL.

% Figures TL and resonators

\subsection{Grounded screening electrode\label{sec:grounded}}
We now analyze the response of the device when the screening electrodes are grounded.
The resulting port admittance is immediately obtained by restricting to the upper $2\times 2$ block of the terminal admittance Eq.   (\ref{eq:admittance-4x4}).\\

Many features of the response can be deduced directly by inspecting the circuit in Fig. \ref{fig:circuit-model} with the terminals $3,4$ connected to ground.
First of all, it is apparent that at the frequencies $\omega_n$ that satisfy the condition in Eq.   (\ref{eq:cond_short}), a full transmission between port 1 and 2 is achieved when the reflection at the circulator is minimized, i.e. when the characteristic impedance of the external circuit matches $1/(2\sigma_{xy})$. 
In this case, the additional delay lines $3,4$ simply add a frequency dependent phase to the transmitted signal; for example, if $Z_{3}(\omega_n)=0$, a voltage wave traveling from port 2 to 1 acquires a phase shift of $\pi$ when it is reflected at the ground, while if $Z_{3}(\omega_n)\rightarrow \infty$, the voltage wave acquires no additional phase since it is reflected at an open.

Voltage waves propagating in asymmetric configurations, e.g.  where the ratio $\tau_{3}/\tau_4$ is chosen appropriately to satisfy simultaneously $Z_{3}(\omega_n)=0$ and $Z_{4}(\omega_n)\rightarrow \infty$ at the frequency $\omega_n$ defined in Eq.   (\ref{eq:cond_short}), pick up an opposite phase depending on the direction they come from. 
This interesting property can be exploited to implement non-reciprocal devices such as the gyrators proposed in \cite{Viola-DiVincenzo, Bosco}. 
In this paper, however, we are mainly interested in reciprocal devices, and thus we restrict our analysis to symmetric setups and we require $\tau_3=\tau_4\equiv\tau_0$, such that the device is reciprocal at any frequency.

\begin{figure}
\includegraphics[width=0.45\textwidth]{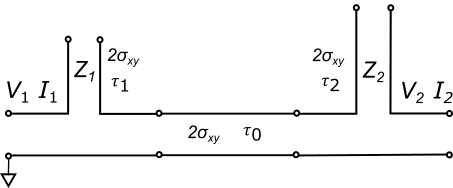}
\caption{Transmission line with a series double-stub tuner.
This circuit is equivalent to a QH droplet with the screening electrodes characterized by the same propagation time $\tau_0$ and connected to ground; the transmission in this region is modeled by a conventional TL with a characteristic impedance $1/(2\sigma_{xy})$.
The capacitive coupling to the driving electrodes is modeled by two stubs (TLs terminated with an open circuit) also having a characteristic impedance of $1/(2\sigma_{xy})$ and a propagation time $\tau_{1,2}$.
}
\label{fig:circuit-grounded}
\end{figure}

For this symmetric device, it is particularly instructive to use conventional microwave techniques to convert the port admittance matrix into a transfer (ABCD) matrix, see e.g. \cite{Pozar}.
The resulting matrix $\mathcal{T}$ can be decomposed into a product of the three transfer matrices that are easy to recognize: 
\begin{equation}
\label{eq:ABCD_matrix_grounded}
\mathcal{T}=\left(
\begin{array}{cc}
 1 & Z_1 \\
 0 & 1
\end{array}
\right)
\left(
\begin{array}{cc}
 \cos \left(\omega  \tau _0\right) & \frac{i \sin \left(\omega  \tau _0\right)}{2 \sigma _{\text{xy}}} \\
 2\sigma _{xy} i \sin \left(\omega  \tau _0\right)  & \cos \left(\omega  \tau _0\right) \\
\end{array}
\right)
\left(
\begin{array}{cc}
 1 & Z_2 \\
 0 & 1
\end{array}
\right).
\end{equation}
For simplicity of notation, we dropped here the explicit dependence on frequency of the impedance $Z_{1,2}$, defined in Eq.   (\ref{eq:terminal-impedance-frequency}), and of the transmission matrix.
This decomposition suggests that the symmetric device acts like a conventional TL cascaded with two `stub tuners' having impedance $Z_i$, as shown in Fig. \ref{fig:circuit-grounded}.
The characteristic impedance of the TL and of the stubs is $1/(2\sigma_{xy})$, while the propagation constants can be found by the equality $\beta_i L_i=\omega \tau_i$.
This circuit is the dual to the parallel double-stub tuner, which is often used in microwave engineering, see e.g. Sec. 5.2 and 5.3 of \cite{Pozar}.\\

We now analyze the main features of this device.
First, by conventional microwave techniques, we can convert the transfer matrix in Eq.   (\ref{eq:ABCD_matrix_grounded}) into the scattering $S$ parameters
\begin{subequations}
\label{eq:s-par-ground}
\begin{align}
   S_{11}&=1-S_{12}\left(\mathcal{T}_{22}+Z_0  \mathcal{T}_{21}\right),\\
   S_{22}&=1-S_{12}\left(\mathcal{T}_{11}+Z_0  \mathcal{T}_{21}\right),\\
   S_{12}=S_{21}&=\frac{2 Z_0  \mathcal{T}_{21}}{\left(\mathcal{T}_{11}+Z_0  \mathcal{T}_{21}\right)\left(\mathcal{T}_{22}+Z_0  \mathcal{T}_{21}\right)-1}.
\end{align}
\end{subequations}
Here, $Z_0$ is the characteristic impedance of the external circuitry and $S_{12}=S_{21}$ because the device is reciprocal.
For simplicity, we also restrict to a more symmetric case and set $\tau_1=\tau_2$; then, $Z_1(\omega)=Z_2(\omega)$, $\mathcal{T}_{11}=\mathcal{T}_{22}$ and $S_{11}=S_{22}$.
The response strongly depends on the impedance matching parameter $\alpha$ and on the ratio of propagation times $p$, defined respectively by 
\begin{subequations}
\begin{flalign}
\alpha&=2\sigma_{xy}Z_0,\\
\label{eq:p-def}
p&=2\frac{\tau_0}{\tau_1}.
\end{flalign}
\end{subequations}
Note that $p$ can be tuned in different ways, for example by modifying the lengths of screening and driving electrodes; in this case, the larger $p$, the longer the TL.\\

\begin{figure}
\includegraphics[width=0.23\textwidth]{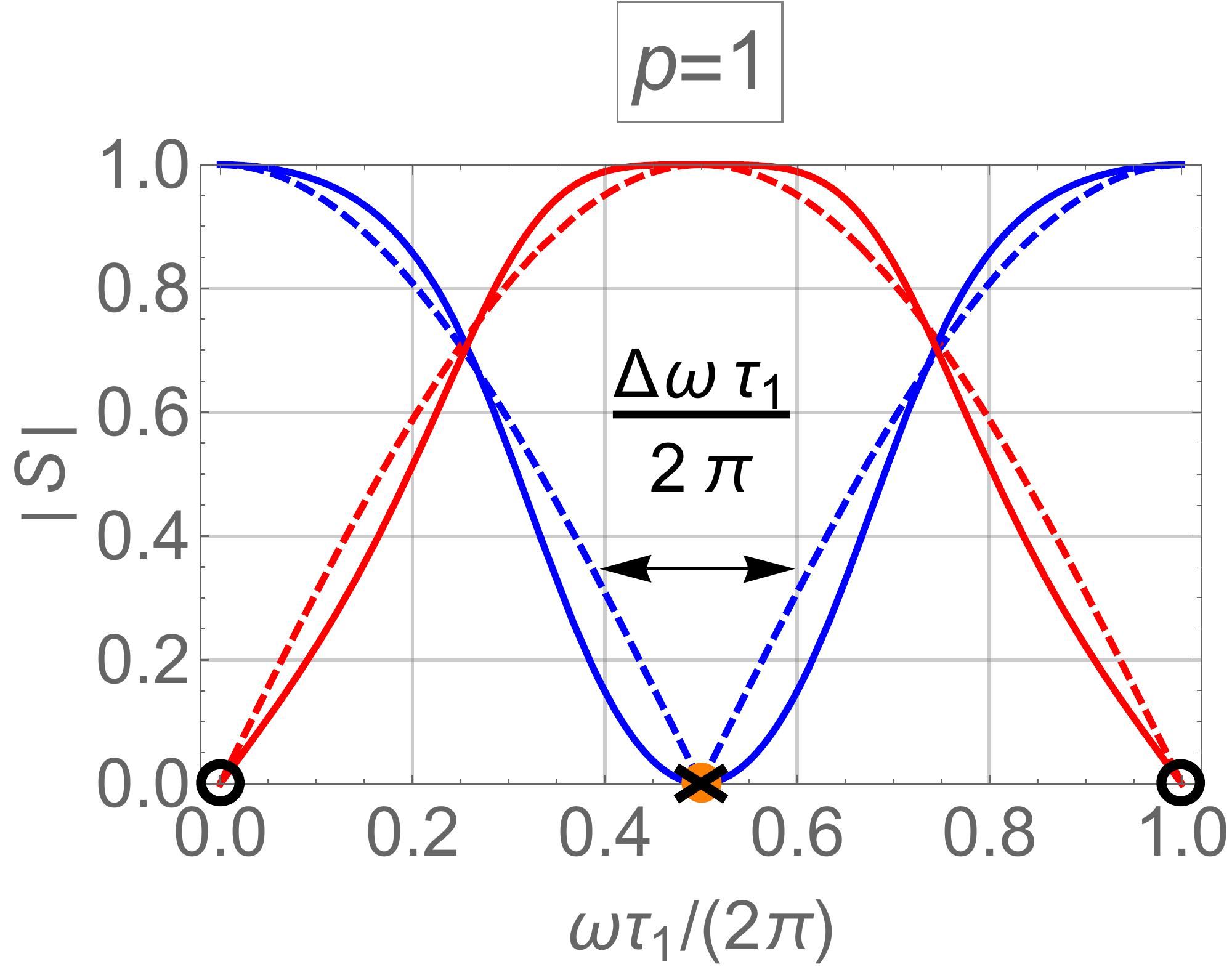}
\includegraphics[width=0.23\textwidth]{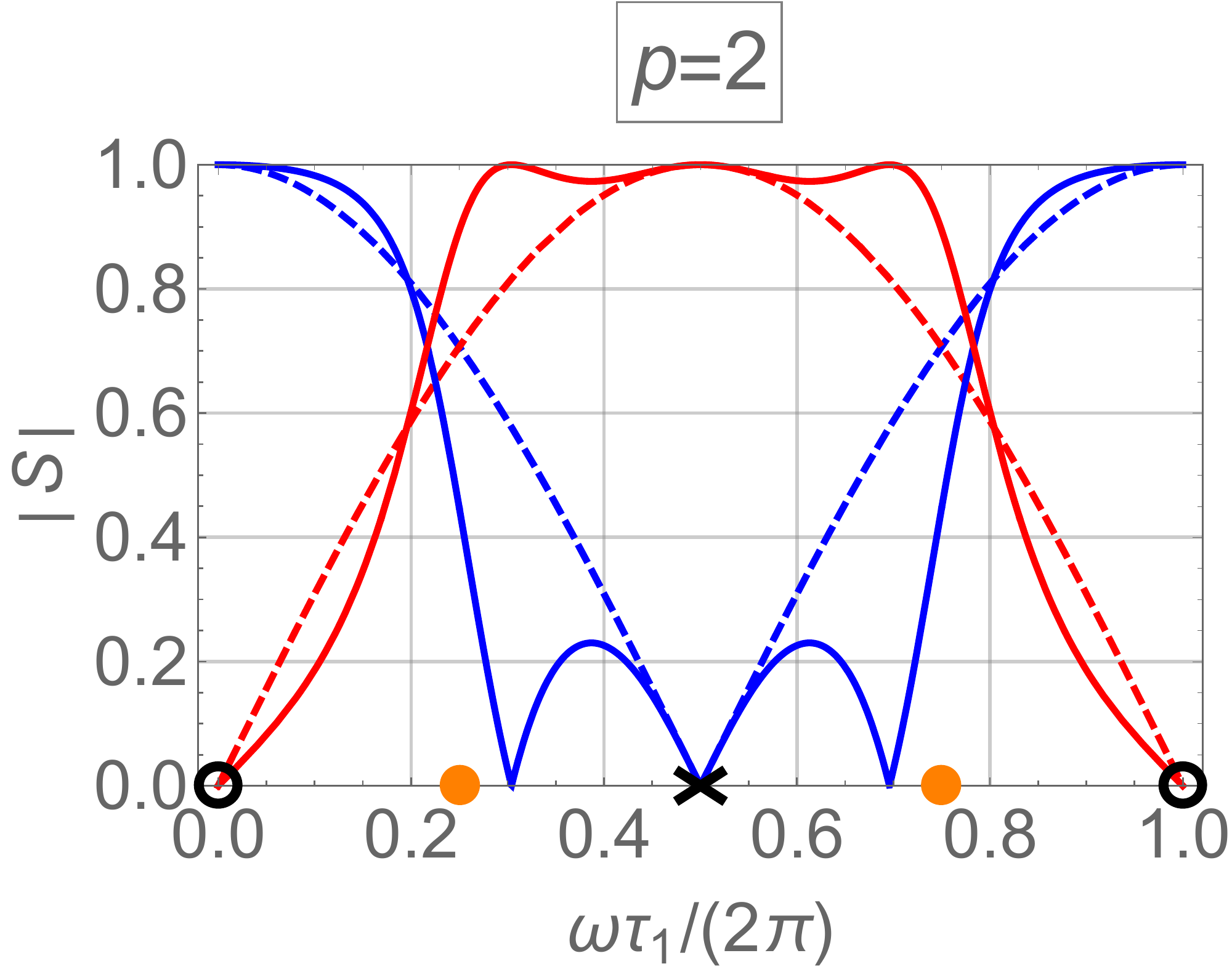}\\
\includegraphics[width=0.23\textwidth]{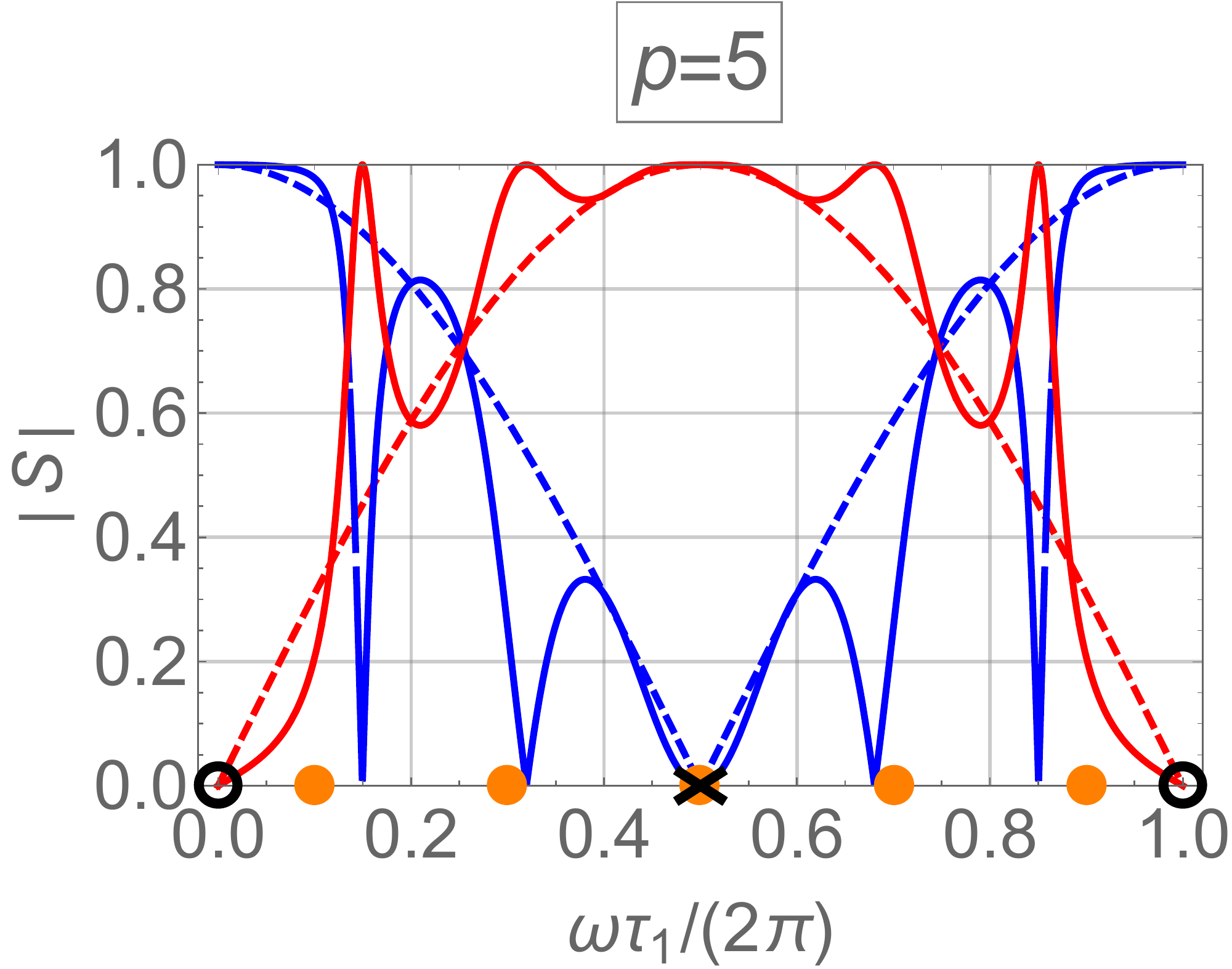}
\includegraphics[width=0.23\textwidth]{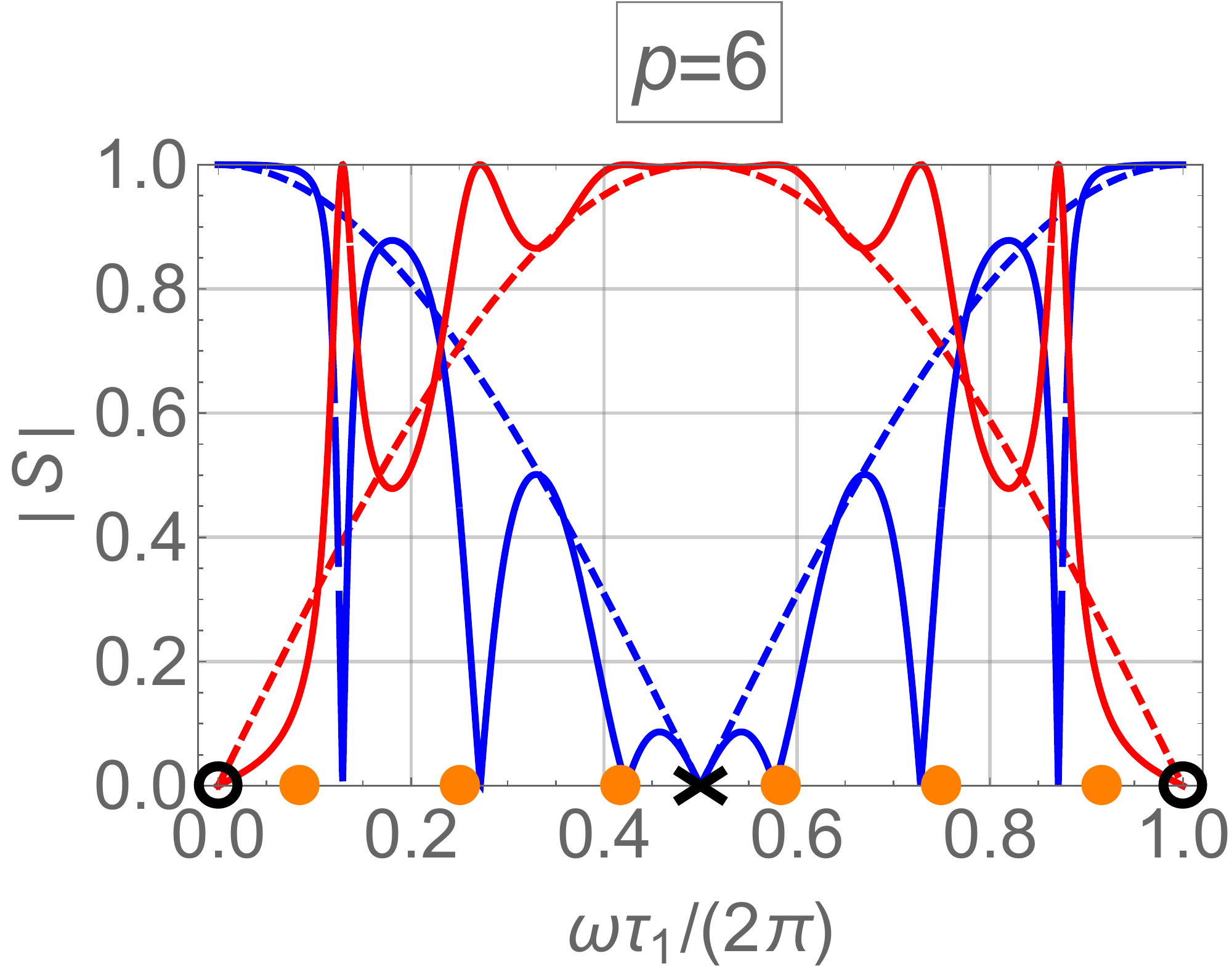}
\caption{Scattering parameters of a QH droplet with grounded and identical screening electrodes.
In the plots, we consider the device to be matched to the external circuitry, i.e. $\alpha= 2\sigma_{xy}Z_0=1$ and we assume a symmetric configuration $\tau_1=\tau_2$.
The different plots are obtained by using different ratios of propagation times  $p=2\tau_0/\tau_1$.
The blue (red) solid line is the absolute value of the scattering parameter associated with reflection (transmission) $\left|S_{11}\right| (\left|S_{12}\right|)$; the dashed lines are the envelope functions that modulate the response obtained by the limit $\tau_0\rightarrow 0$; they capture the response of a configuration with floating screening electrodes, as described in Sec. \ref{sec:floating-gates}. 
The envelope functions of $\left|S_{11}\right|$ vanish at the central resonances   $\omega_n$  (black crosses) defined in Eq. (\ref{eq:cond_short}), and they attain the maximum value of one at the frequencies $\omega_m$ (black circles) defined in Eq. (\ref{eq:cond_open}).
The fast oscillations depend on the transmission line and they are associated with the resonances at the frequencies $\omega_l$ (orange dots) defined in Eq. (\ref{eq:Tl-resonances}).
  }
\label{fig:scattering_grounded}
\end{figure}

In Fig. \ref{fig:scattering_grounded}, we show the frequency dependence of the absolute value of the scattering parameters when $\alpha=1$.
For convenience, we restrict our analysis to integer values of $p$, because in this case the $S$ parameters are periodic in frequency with period $\omega=2\pi/\tau_1$; we stress, however, that perfect transmission can be achieved for any real value of $p$.
The $S$ parameters are characterized by fast oscillations modulated by a smooth envelope function.
This envelope function is attained by taking the limit $\tau_0\rightarrow 0$, which corresponds to an infinitesimal length of the screening electrodes and therefore captures the response of the two equal stubs with impedance $Z_1(\omega)$. 

Also, as expected, the envelope function of $\left|S_{11}\right|$ drops to zero at the frequencies $\omega_n$ for which the condition in Eq.   (\ref{eq:cond_short}) is satisfied, while it reaches the maximum value of one at the frequencies $\omega_m$ when the opposite condition (\ref{eq:cond_open}) holds. This indicates that at these frequencies the device is respectively perfectly transmitting and perfectly reflective. In Fig. \ref{fig:scattering_grounded}, we mark $\omega_n$ and $\omega_m$ with black crosses and black circles, respectively; we will refer to the frequencies $\omega_n$ defined in Eq.  (\ref{eq:cond_short}) as the central resonances.

The fast oscillations are caused by the transmission line.
In particular, they are associated with the resonances occurring at the frequencies 
\begin{equation}
\label{eq:Tl-resonances}
\omega_l=\frac{\pi}{\tau_0}\left(\frac{1}{2}+l\right)=\frac{2\pi}{p\tau_1}\left(\frac{1}{2}+l\right),
\end{equation}
at which the input impedance of the transmission line segment would vanish if it was isolated with open boundary conditions; here $l\in \mathbb{N}_0$. The frequencies $\omega_l$ are marked with orange dots in Fig. \ref{fig:scattering_grounded}. 
%In particular, a transmission line segment isolated and terminated with open boundary conditions has an input impedance $\propto \cot(\omega\tau_0)$ \cite{Pozar}. The input impedance would . 
% obtained when the input impedance of the the transmission line segment terminate is a short. This happens at frequencies egment is a short. This happens at frequencies  ssociated with the resonances of transmission line, that occur at the frequencies $\omega=\pi(1/2+l)/\tau_0$. In this case at which the input impedance of the 
%These resonances are obtain  the transmission line segment were isolated (i.e., a cavity) with open (?) boundary conditions.occur at the frequencies at which the input impedance of the transmission line (terminated by an open circuit) is  to the frequencies at which the input impedance of the  that would occur, without the two stubs, at frequencies  
There are exactly $\lceil p\rceil$ of such resonances per period of the envelope function ($\lceil x\rceil$ indicates the ceiling of $x$).
Note that away from $\omega_n$, the exact frequency of these fast resonances differs from $\omega_l$ because of the influence of the delay lines $Z_1(\omega)$. 
Interestingly, the response at the central resonances $\omega_n$ presents qualitatively different behavior depending on the parity of $p$. 
When $p$ is even, the transmission coefficient is real $S_{12}=(-1)^{p/2}$ and the reflection increases linearly in $\omega$, with the same slope as the smooth modulating function.
On the other hand, when $p$ is odd, one of the fast TL resonances occurs exactly at frequency $\omega_n$.
In this case, the transmitted signal acquires an imaginary phase $S_{12}=(-i)^{p}$ and additionally $\omega_n$ becomes a sweet spot at which both $S_{11}$ and its first derivative in frequency vanish,  increasing the bandwidth of the device.
Varying $p$ as a real parameter from odd to even, one observes a transition between the two situations: close to $\omega_n$, the reflection parameter, to linear order in frequency, simplifies into  $S_{11}\approx i \left(1+e^{-i \pi  p}\right) (\omega-\omega_n)\tau_1/4$, and thus the slope of $|S_{11}|$ continuously oscillates as a function of $p$ from the minimal value $0$, when $p$ is odd, to the maximal value of $\tau_1/2$, when $p$ is even.
Since we are mostly interested in TLs where the screening electrodes are much longer than the driving ones, we will now focus on the large $p$ limit, where the difference between the two situations is small.
Note however that the parameter $p$ depends also on the ratio of EMP velocities $v_i$, given in Eq.   (\ref{eq:EMP-vel-app}), in the screening and driving regions, and therefore it can be tuned also by modifying the capacitive coupling to the corresponding electrodes, for example, by bringing the electrodes closer to the quantum Hall edge in one region.
Also, more importantly, we anticipate that the parity of $p$ has interesting consequences when more QH droplets are cascaded, as described in Sec. \ref{sec:meta-material-tl}.\\

When $p$ is large, the bandwidth of the device becomes independent of $p$ and can be estimated from the slope of the smooth function modulating $|S_{11}|$. 
Expanding this  function to linear order in frequency in the vicinity of $\omega_n$, we find
\begin{equation}
\label{eq:bandwidth}
\frac{\Delta\omega}{\omega_n}=\frac{4}{\pi(1+2n)}\left|S_{11}^{\mathrm{max}}\right|.
\end{equation}
Here, $\Delta\omega$ is the broadening of the envelope function of $\left|S_{11}\right|$ close to $\omega_n$, as shown in Fig. \ref{fig:scattering_grounded}. 
The linear approximation for the bandwidth is quite accurate, giving an error below $3\%$ up to relative high reflection $\left|S_{11}\right|\lesssim 0.4$.\\

\begin{figure}
\includegraphics[width=0.23\textwidth]{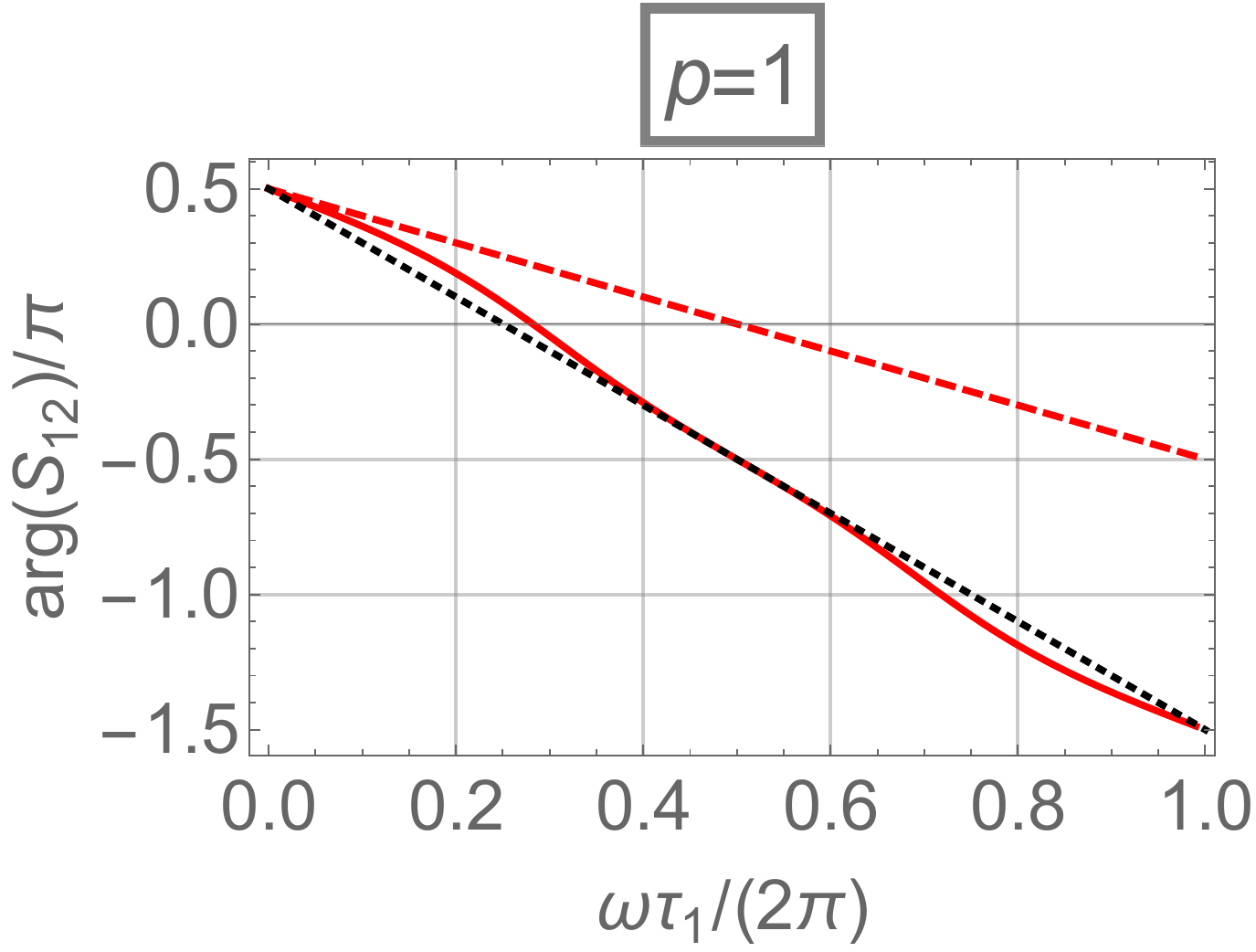}
\includegraphics[width=0.23\textwidth]{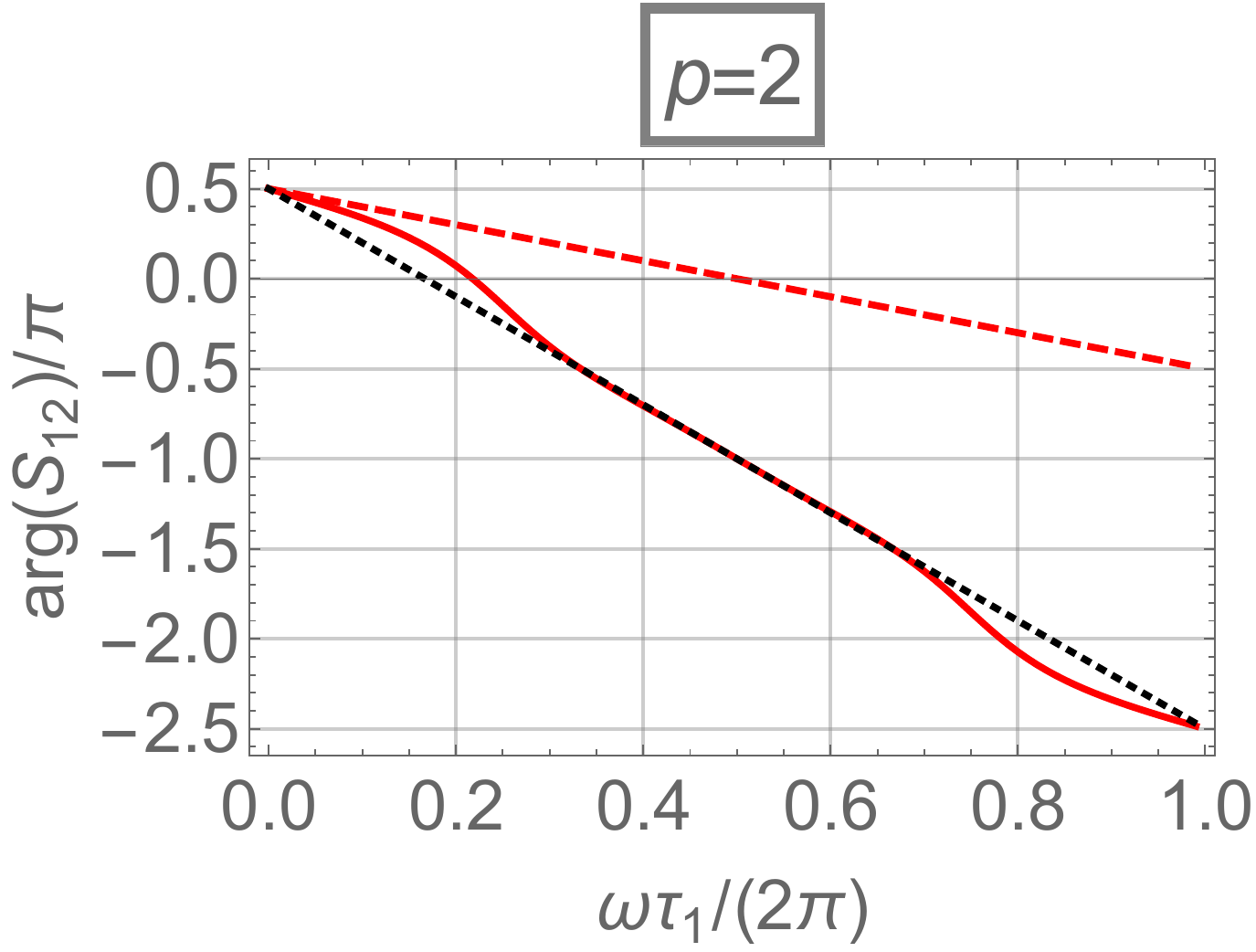}\\
\includegraphics[width=0.23\textwidth]{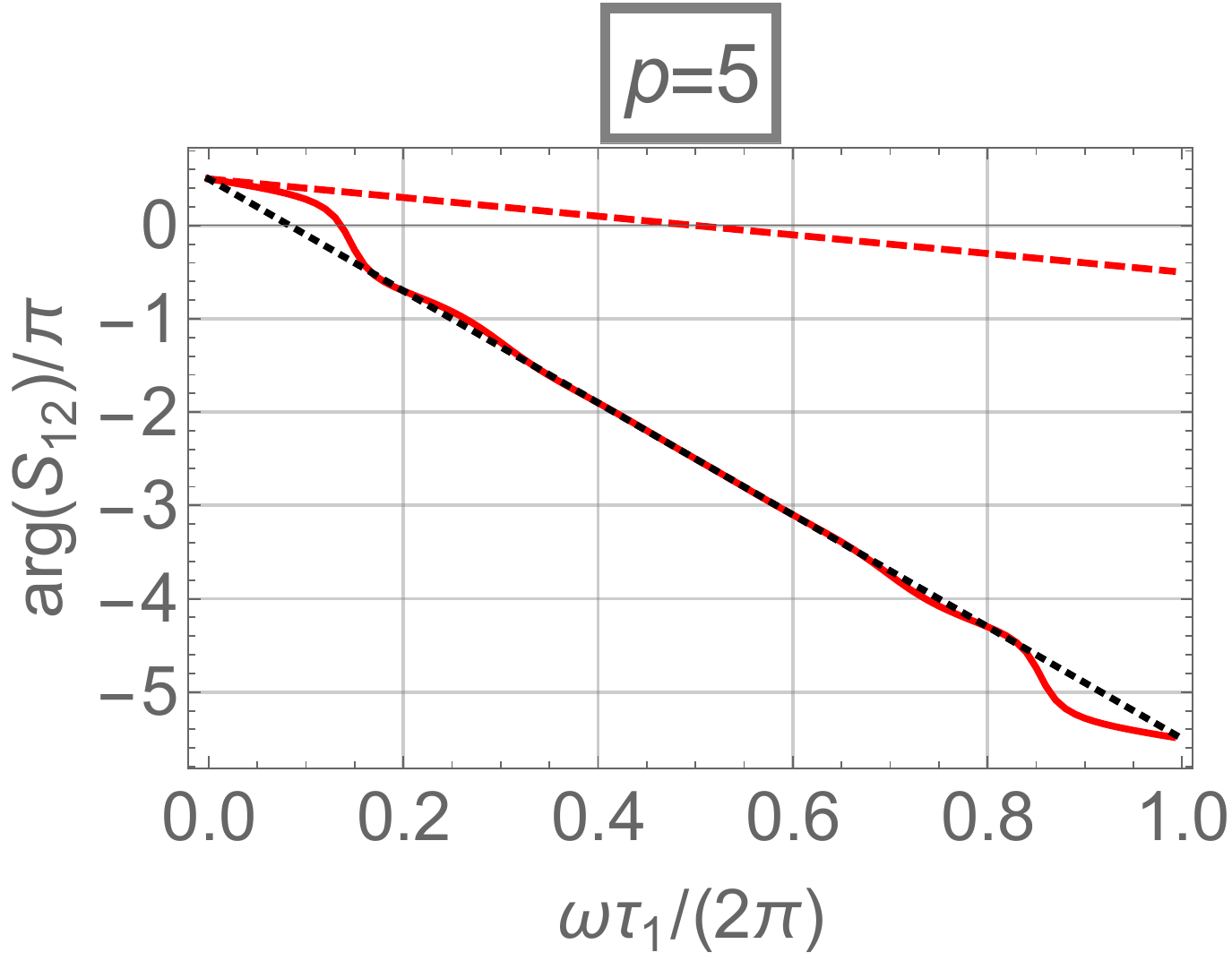}
\includegraphics[width=0.23\textwidth]{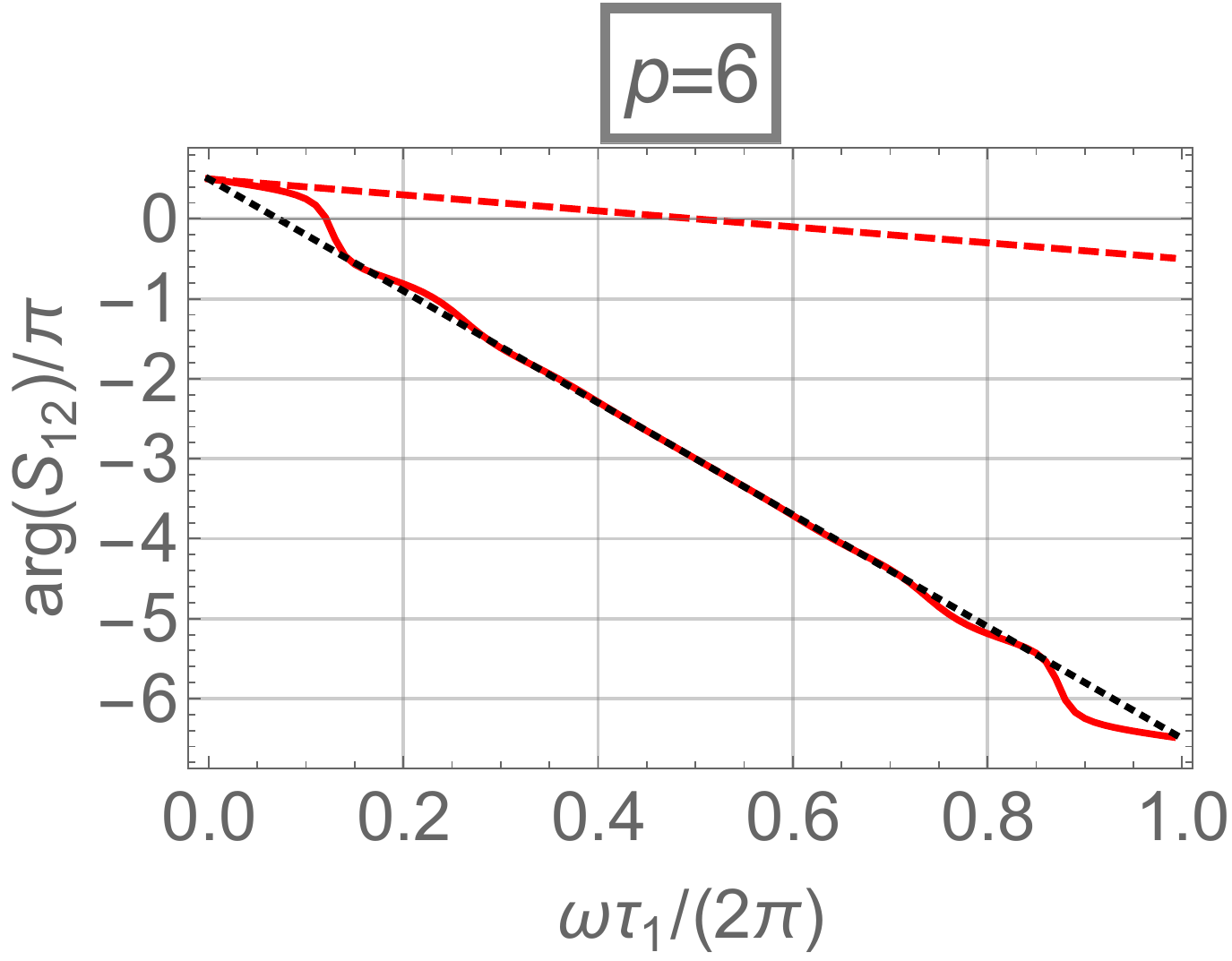}
\caption{Phase of the transmission parameter $S_{12}$.
The device is matched to the external circuitry, i.e. $\alpha=1$, the screening electrodes are identical, i.e. $\tau_1=\tau_2$ and the different plots are obtained by using different ratios of propagation times  $p$.
The solid red lines show the phase of the scattering parameter  of a QH droplet with grounded screening electrodes, while the red dashed lines show the phases of the envelope function which modulates the response, and they are obtained by taking the limit $\tau_0\rightarrow 0$; these dashed lines capture the response of the setup described in Sec. \ref{sec:floating-gates}. 
The dotted black lines are obtained by using the linear expansion of the phase in Eq. (\ref{eq:phase-ground}). 
\label{fig:phase_scattering_grounded}}
\end{figure}

In Fig. \ref{fig:phase_scattering_grounded}, we show the phase of the scattering parameter $S_{12}$ associated with the transmission for different values of $p$ in a matched setup with $\alpha=1$. Close to the central resonance peaks, the phase of $S_{12}$ is almost linear in the first period of the envelope function and is well approximated by 
\begin{equation}
\label{eq:phase-ground}
\frac{\arg(S_{12})}{\pi}\approx \frac{1}{2}-(1+p)\frac{\omega \tau_1}{2\pi}.
\end{equation}

Eq. (\ref{eq:phase-ground}) is a sum of two contributions: the phase accumulated in a conventional transmission line and the phase accumulated in the two stubs, i.e. $\arg(S_{12})=\arg(S_{12}^{\text{TL}})+2\arg(S_{12}^{\text{ST}})$.
In particular, a matched TL, characterized by a propagation time $\tau_0=p\tau_1/2$, has an off-diagonal scattering matrix with $S_{12}^{\text{TL}}=e^{-i \omega\tau_0}$; thus $\arg(S_{12}^{\text{TL}})=- p \omega\tau_1/2$.
The phase accumulated in each stub, close to the central resonances, where $S_{ii}\approx 0$, is given by $\arg(S_{12}^{\text{ST}})\approx(\pi-\omega \tau_1)/4$; the phase of the envelope function is twice this value, i.e. $2\arg(S_{12}^{\text{ST}})$.
Note that  a phase that changes linearly in frequency  implies that the device is not dispersive and thus the transmitted voltage wave preserves the original shape, with a delay time of $(1+p)\tau_1/2$, as one can see by an inverse Fourier transform of $V^{\text{out}}(\omega)=S_{12}(\omega) V^{\text{in}}(\omega)$.\\

To conclude our analysis, we mention here that adding two additional stubs in series with a TL is a well-known procedure to match different impedances \cite{Pozar}; this means that by appropriately choosing the propagation times $\tau_1$ and $\tau_2$, the input impedance $Z_{\text{in}}(\omega)$ of this device can be tuned to match different loads at the desired frequency of operation.
In particular, if we terminate port 2 with a load $Z_L(\omega)$ to ground, it is straightforward to verify that the input impedance seen at port 1 is:
\begin{equation}
\label{eq:input-output-impedance}
z_{\text{in}}(\omega)=\frac{\csc^2(\omega \tau_0)}{z_L(\omega)+z_2(\omega)-i\cot (\omega \tau_0)}-i\cot (\omega \tau_0)+ z_1(\omega),
\end{equation}
where, for simplicity of notation, we introduced the dimensionless impedances $ z_j(\omega)\equiv 2\sigma_{xy}Z_j(\omega)$.
%To find the range of load impedances that can be matched, we separate the real and imaginary part of load and input impedance $z_j=z'_j+iz''_j$ and, considering that $z_{1,2}=i z_{1,2}'' $ are purely imaginary, we get from the real part of Eq.   (\ref{eq:input-output-impedance}) the condition on the real part of the load
Decomposing these impedances into their real and imaginary part  $z_j=z'_j+iz''_j$, we obtain from Eq.   (\ref{eq:input-output-impedance}) the following conditions on $z''_{1,2}$
\begin{subequations}
\label{eq:z_impedance-matching}
\begin{flalign}
z''_{1}(\omega)&=z''_{\text{in}}(\omega) +\cot( \omega\tau_0)\pm\sqrt{ \frac{z'_{\text{in}}}{z'_{L}} (\csc^2(\omega \tau_0)-z'_L z'_{\text{in}})} , \\
z''_{2}(\omega)&=-z''_{L}(\omega)+\cot( \omega\tau_0)\pm\sqrt{ \frac{z'_{L}}{z'_{\text{in}}} (\csc^2(\omega \tau_0)-z'_L z'_{\text{in}})} .
\end{flalign}
\end{subequations}

Therefore, when $\tau_0$ is fixed, the real part of input and load impedances are bounded by the inequality
\begin{equation}
\label{eq:matching-condition}
0\leq z'_L(\omega) z'_{\text{in}}(\omega)\leq \csc^2(\omega \tau_0),
\end{equation}
which determines the range of load impedances that can be matched with a given input impedance by this circuit.

For example, this device can be used to match a real load of the order of the quantum of resistance $z_L'\approx 1$ to a much smaller real input  $z_{\text{in}}=z'_{\text{in}}\ll 1$, provided that the propagation times  $\tau_{0,1,2}$ are chosen to satisfy the inequality in Eq.   (\ref{eq:matching-condition}) and the equalities in Eq.   (\ref{eq:z_impedance-matching}) at the frequency of operation.

%Expanding these impedances into their real and imaginary part  $z_j=z'_j+iz''_j$, we obtain from the real part of Eq.   (\ref{eq:input-output-impedance}) the following condition on $z'_L$
%\begin{equation}
%z'_L(\omega)=\frac{\csc^2(\omega \tau_0)}{2 z'_{in}(\omega)}\left(1\pm \sqrt{1-f(\omega)^2}\right),
%\end{equation}
%with
%\begin{equation}
%f(\omega)=2 z_{in}'(\omega)\sin(\omega\tau_0)^2\left(z''_{L}(\omega)+z_2''(\omega)-\cot(\omega \tau_0)\right).
%\end{equation}
%Therefore, when $\tau_0$ is fixed, the real part of input and load impedances are bounded by the inequality
%\begin{equation}
%\label{eq:matching-condition}
%0\leq z'_L(\omega) z'_{in}(\omega)\leq \csc^2(\omega \tau_0),
%\end{equation}
%which determines the range of load impedances that can be matched.
%This means, for example, that a real load $z_L=z'_L$ can only be matched with a real input $z_{in}=z'_{in}$ in the range defined by Eq.   (\ref{eq:matching-condition}); in this case, the propagation times $\tau_1$ and $\tau_2$ need to be chosen to satisfy at the frequency of operation the equality 
%\begin{equation}
%z_1''=z_2''=\cot( \omega\tau_0)\pm\sqrt{ \frac{z'_{in}}{z'_{L}} (\csc^2(\omega \tau_0)-z'_L z'_{in})}.
%\end{equation}

%Figures

\subsection{Floating screening electrodes\label{sec:floating-gates}}
Here, we study what happens when the two screening electrodes are left floating.
In this case, the port admittance matrix can be derived from the terminal admittance in Eq.   (\ref{eq:admittance-4x4}) by  first computing the values of screening potentials $V_{3,4}$ (as a function of the driving potentials $V_{1,2}$) that guarantee $I_3=I_4=0$ and then using these results to construct the $2\times 2$ port admittance matrix.
Proceeding as before, we convert this port admittance into a transfer matrix and we find that the resulting ABCD matrix exactly corresponds to the limit $\tau_0\rightarrow 0$ of Eq.   (\ref{eq:ABCD_matrix_grounded}).

The circuit model in Fig. \ref{fig:circuit-grounded} is then modified by considering  only the series of the two delay lines with impedance $Z_1(\omega)$ and $Z_2(\omega)$.
This result can be understood by observing that when the screening electrodes are left floating, no net current flows into them, and thus they cannot contribute to the response of the device. For this reason, in contrast to the case described in Sec. \ref{sec:grounded}, this device is always reciprocal for any value of $\tau_{3,4}$. 
Comparing with \cite{Viola-DiVincenzo}, one realizes that this boundary condition also mimics the response of the device when the regions $3$ and $4$ are not coupled to screening electrodes.

Also, the scattering parameters of this device are exactly given by the envelope functions discussed in Sec. \ref{sec:grounded} and whose absolute values are shown with dashed lines in Fig. \ref{fig:scattering_grounded}; at the central resonances, the transmitted and incident voltage waves have always the same phase.

%Finally, by taking the limit $\tau_0\rightarrow 0$ in Eq.   (\ref{eq:matching-condition}), one can see that this configuration allows to match any real load to any real input impedance.
Finally, by taking the limit $\tau_0\rightarrow 0$ in Eq.   (\ref{eq:input-output-impedance}), one gets
\begin{equation}
z_{in}(\omega)=z_L(\omega)+z_1(\omega)+z_2(\omega),
\end{equation}
from which it follows that the real part of the input impedance is always equal to the real part of the load impedance, and thus this construction cannot be used for impedance matching.

\subsection{Parasitics\label{sec:parasitics}}
\begin{figure}
\includegraphics[scale=0.30]{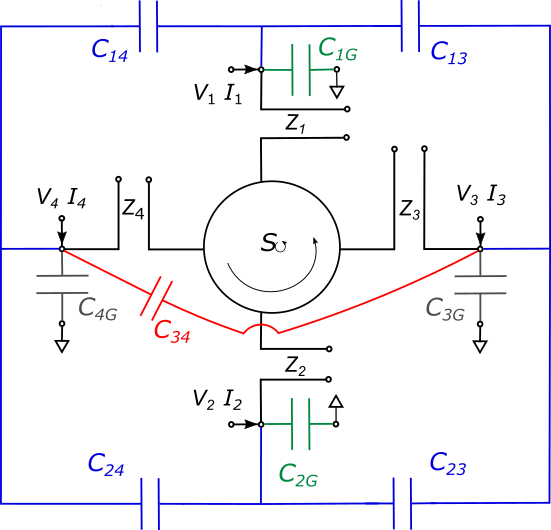}
\caption{Equivalent circuit model including parasitic capacitors.
The circuit equivalent of the QH droplet in Fig. \ref{fig:circuit-model} is augmented by a parallel circuit that include the possible capacitive coupling between external electrodes and to ground; we do not include the coupling $C_{12}$ between the driving terminals because it is negligible for long TLs.
Some of these additional capacitors modify the response in a qualitatively similar manner and they are thus drawn with the same color. Following the notation of the main text the gray, red, green and blue capacitors are respectively associated with the charging times $T_0$, $T_1$, $T_2$ and $T_3$.}
\label{fig:circuit-parasitics}
\end{figure}

The capacitive coupling between adjacent electrodes is known to strongly influence the response of QH circulators and gyrators \cite{Reilly,Placke,Bosco}.
For this reason, we now analyze how the TL is affected by these parasitics.
We consider the augmented circuit model shown Fig. \ref{fig:circuit-parasitics}, where we included all the possible parasitic couplings, except for the one that directly connects the driving terminal $1$ and $2$; this contribution is negligible in the limit of long screening electrodes.

The network of parasitic capacitors is in parallel to the transmission line and      so the  terminal admittance in Eq.    (\ref{eq:admittance-4x4}) modifies as
\begin{equation}
Y\rightarrow Y+Y_{P},
\end{equation}
with
\begin{subequations}
\begin{flalign}
(Y_P)_{ii}&=i\omega \sum_{j\neq i} C_{ij}+i\omega C_{iG},\\
(Y_P)_{ij}&=-i\omega C_{ij}.
\end{flalign}
\end{subequations}

The 2-port admittance is obtained from this augmented admittance using the same procedure described in Sec. \ref{sec:floating-gates}, and from the resulting matrix we straightforwardly obtain the $S$-parameters. We assume again that the characteristic impedances of the external circuit $Z_0$ and of the QH device are matched, i.e. $\alpha=1$.\\

We begin our analysis by focusing only on the capacitances that connect the screening electrodes to ground $C_{3G},C_{4G}$.
For simplicity, we take them to be equal and we introduce the timescale $T_0\equiv C_{3G}/(2\sigma_{xy})$, that characterizes the charging time of these capacitors.

The effect of $T_0$ can be understood by realizing that in the limiting cases, $T_0\rightarrow \infty$ and $T_0\rightarrow 0$, the response of the device reduces to the one described in Sec. \ref{sec:grounded} and \ref{sec:floating-gates}, respectively.
For finite values of $T_0$ the $S$-parameters interpolate between the ones of the two configurations; in particular, for small $T_0$, Fano-like resonances begin to appear on top of the envelope function and, as $T_0$ is increased, they smoothly evolve to the features shown in Fig. \ref{fig:scattering_grounded}.
Also, for finite $T_0$, the response is not periodic in frequency anymore: in particular, in the high frequency limit $\omega T_0\gg1$ the screening electrodes always act like they are shorted to ground.

We divide the remaining parasitic capacitors into a few groups exhibiting a qualitatively similar behavior. We study them separately and we show the differences in the limits $T_0\rightarrow 0$ and $T_0\rightarrow\infty$, that correspond to floating and grounded screening electrodes, respectively. These groups of capacitors are shown with different colors in Fig. \ref{fig:circuit-parasitics}.\\

\begin{figure}
\includegraphics[width=0.23\textwidth]{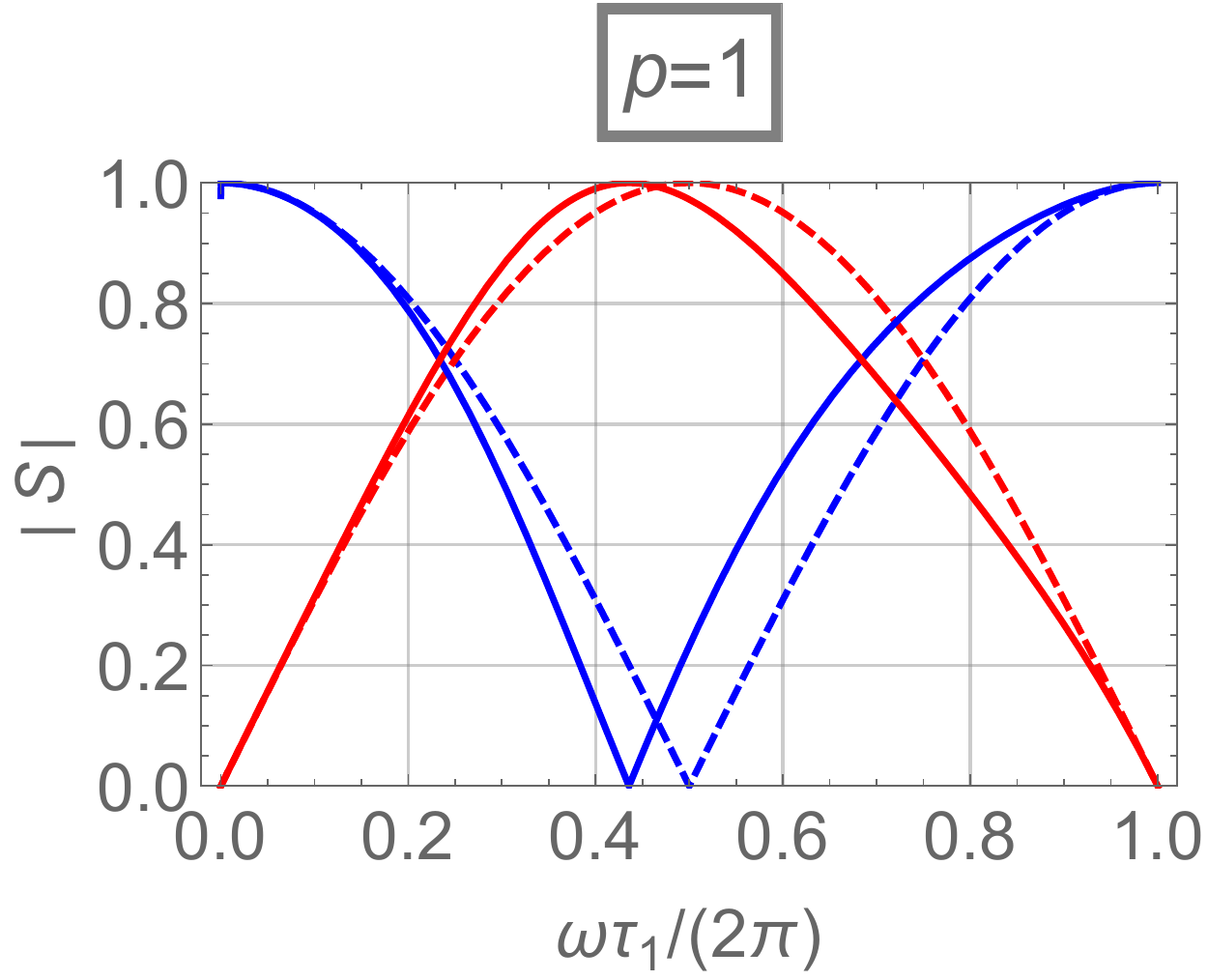}
\includegraphics[width=0.23\textwidth]{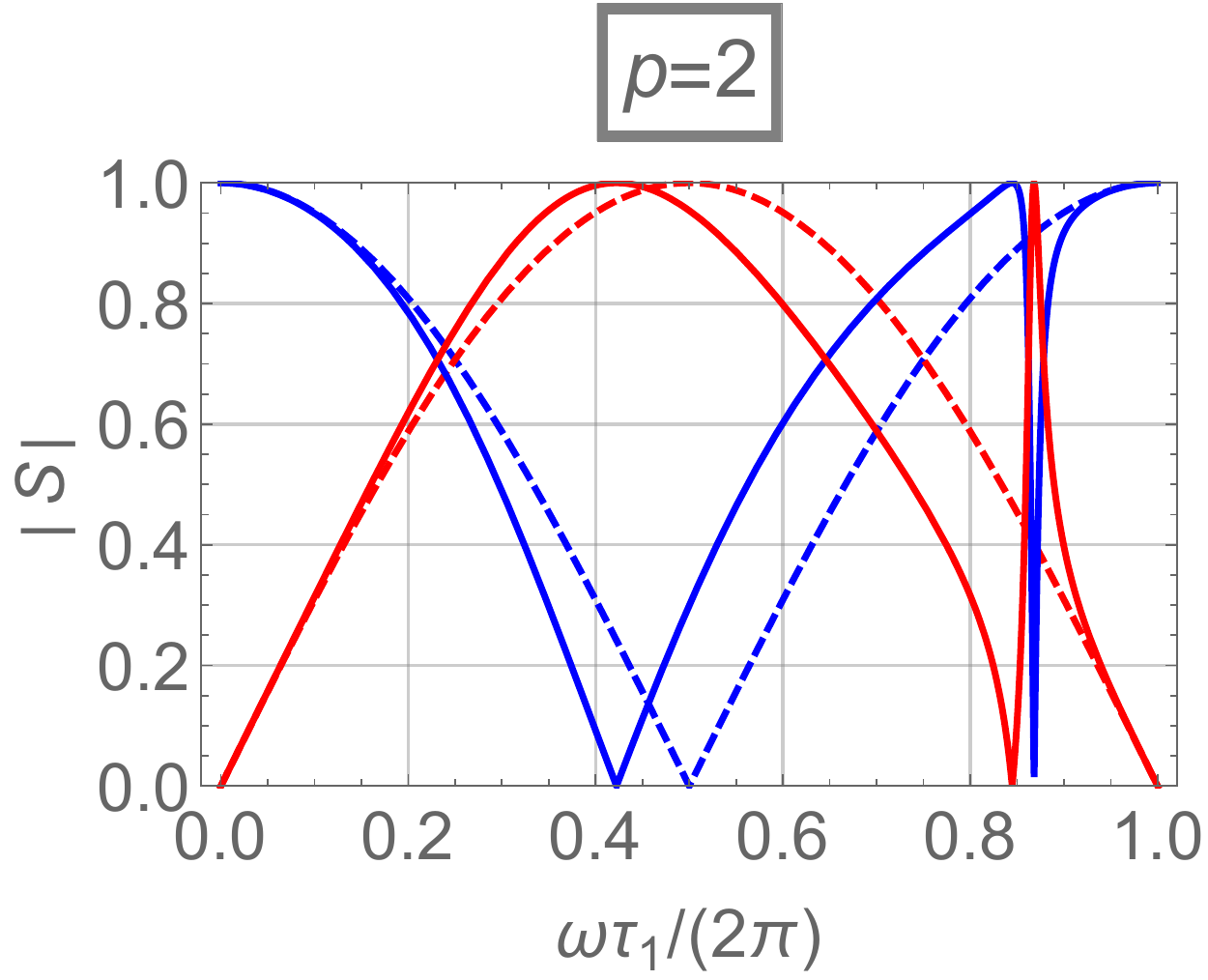}\\
\includegraphics[width=0.23\textwidth]{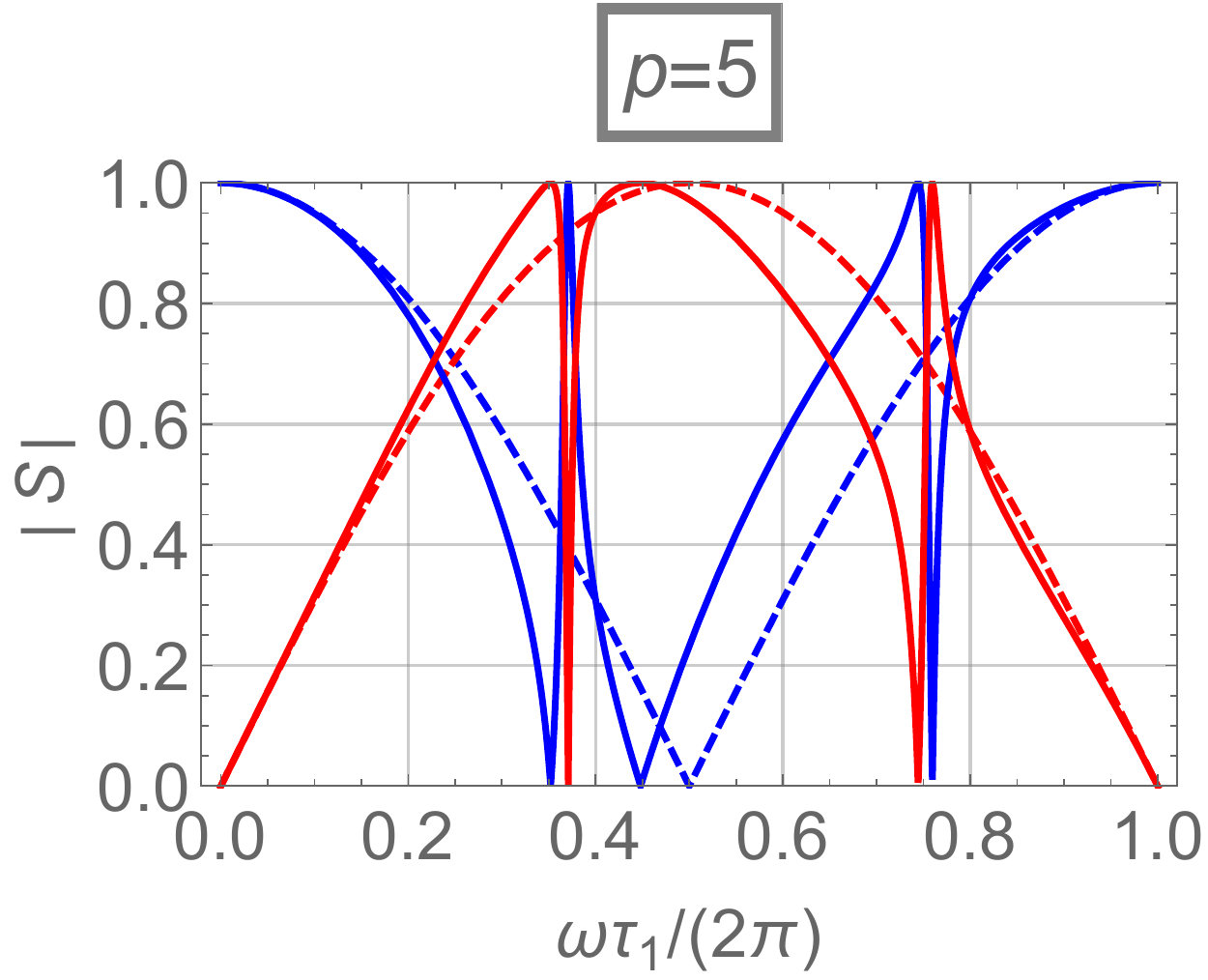}
\includegraphics[width=0.23\textwidth]{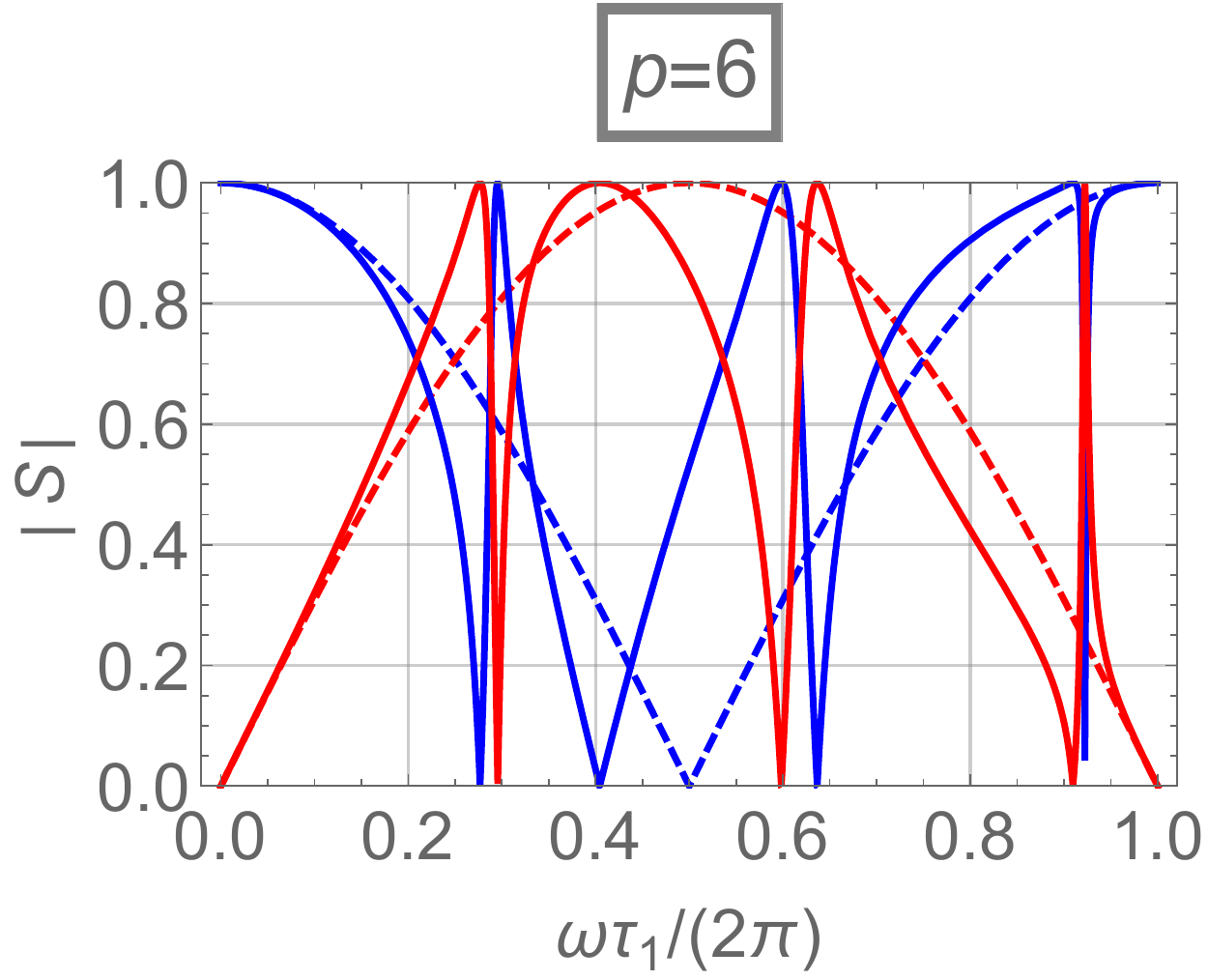}
\caption{Changes in the scattering parameters of a QH droplet due to a small value of the parasitic capacitor $C_{34}\equiv 2\sigma_{xy}T_1$.
In the plots, the screening gates are assumed to be floating, i.e. $T_0=0$,  and we consider the device to be matched with the external circuitry $\alpha\equiv 2 \sigma_{xy}Z_0=1$ and symmetric, i.e. $\tau_1=\tau_2$ and $\tau_3=\tau_4=\tau_1 p/2$.
The different plots show the scattering parameters for different values of $p$.
The blue (red) solid line represents the absolute value of $\left|S_{11}\right|$ $(\left|S_{12}\right|)$ when $T_1/\tau_1=0.05$, while the dashed lines show the response in the same configuration in the case without parasitics.}
\label{fig:scattering_floating_par34}
\end{figure}

We focus now on the capacitor $C_{34}$ which connects opposite screening electrodes and we introduce the charging time $T_1\equiv C_{34}/(2\sigma_{xy})$.
Clearly $T_1$ does not affect the response of the device when $\omega T_0\gg 1$, but it strongly influences the performance in the opposite limit.
Fig. \ref{fig:scattering_floating_par34} shows how the absolute value of the $S$ parameters change from the ideal case when the screening electrodes are left floating ($T_0=0$) for a small variation of $T_1/\tau_1=0.05$ and for different values of the ratio of propagation time $p$, defined in Eq. (\ref{eq:p-def}).

First, note that the central resonances, defined without parasitics by Eq.    (\ref{eq:cond_short}), are shifted to lower frequency and that the corresponding bandwidth decreases; for small enough $T_1/\tau_1$, the shifted resonance frequencies $\tilde{\omega}_n$ are approximately given by
\begin{equation}
\label{eq:shift-resonance-frequency-34}
\frac{(\tilde{\omega}_n-\omega_n) \tau_1}{4\pi(1+2n)}\approx-\frac{T_1}{\tau_1}+4\left(1+\pi(-1)^n\left(\frac{1}{2}+n\right)\right)\left(\frac{T_1}{\tau_1}\right)^2.
\end{equation}
%A finite $T_0$ shifts back these resonances to the frequencies $\omega_n$ and restores the bandwidth of Eq.    \ref{eq:bandwidth}.
In contrast, the resonances $\omega_m$ associated with perfect reflections, defined without parasitcs by Eq.    (\ref{eq:cond_open}), are not affected by $T_1$: when the driving impedances act as open circuits, voltage waves are always reflected back, regardless of how the screening electrodes are connected. 

Also, for a finite $T_1$, the screening electrodes begin to affect the response because of the finite net current flowing into them.
In particular, when $p>1$, the scattering parameters present additional Fano-like resonances, that are associated with the limit $Z_{3,4}(\omega)\rightarrow\infty$, when the screening electrodes act like open circuits; there are $\lceil (p-1)/2\rceil$ of such resonances in the first period. 
%As $T_1$ increases, the exact resonances are shifted to lower frequency, but in contrast to what happens to the central resonance, their broadening increases.
As $T_1$ increases, these peaks are shifted to lower frequency, but in contrast to what happens to the central resonance, their broadening increases.

Note that these resonances are qualitatively different from the ones examined in Sec. \ref{sec:grounded}; in fact, they are not modulated by a smoother function, but their amplitudes range from the minimal value of 0 to the maximal value of 1.
For this reason, when the screening electrodes are left floating, a finite $T_1$ strongly degrades the performances of long devices with $p\gg 1$.
In this case, coupling the screening electrodes to ground by a finite $T_0$ provides some advantages. 
In fact, when $T_0$ increases, the $S$ parameters gradually return the structure shown in Fig. \ref{fig:scattering_grounded}: the resonances $\tilde{\omega}_n$ are shifted back to $\omega_n$ and their bandwidth is restored, the amplitudes of the additional Fano resonances recover the original modulation; also, the $\lceil p/2\rceil$ peaks associated with the limit $Z_{3,4}(\omega)\rightarrow 0$ appear and all the $\lceil p\rceil$ resonances discussed in Sec. \ref{sec:grounded} are  recovered.

\begin{figure}
\includegraphics[width=0.23\textwidth]{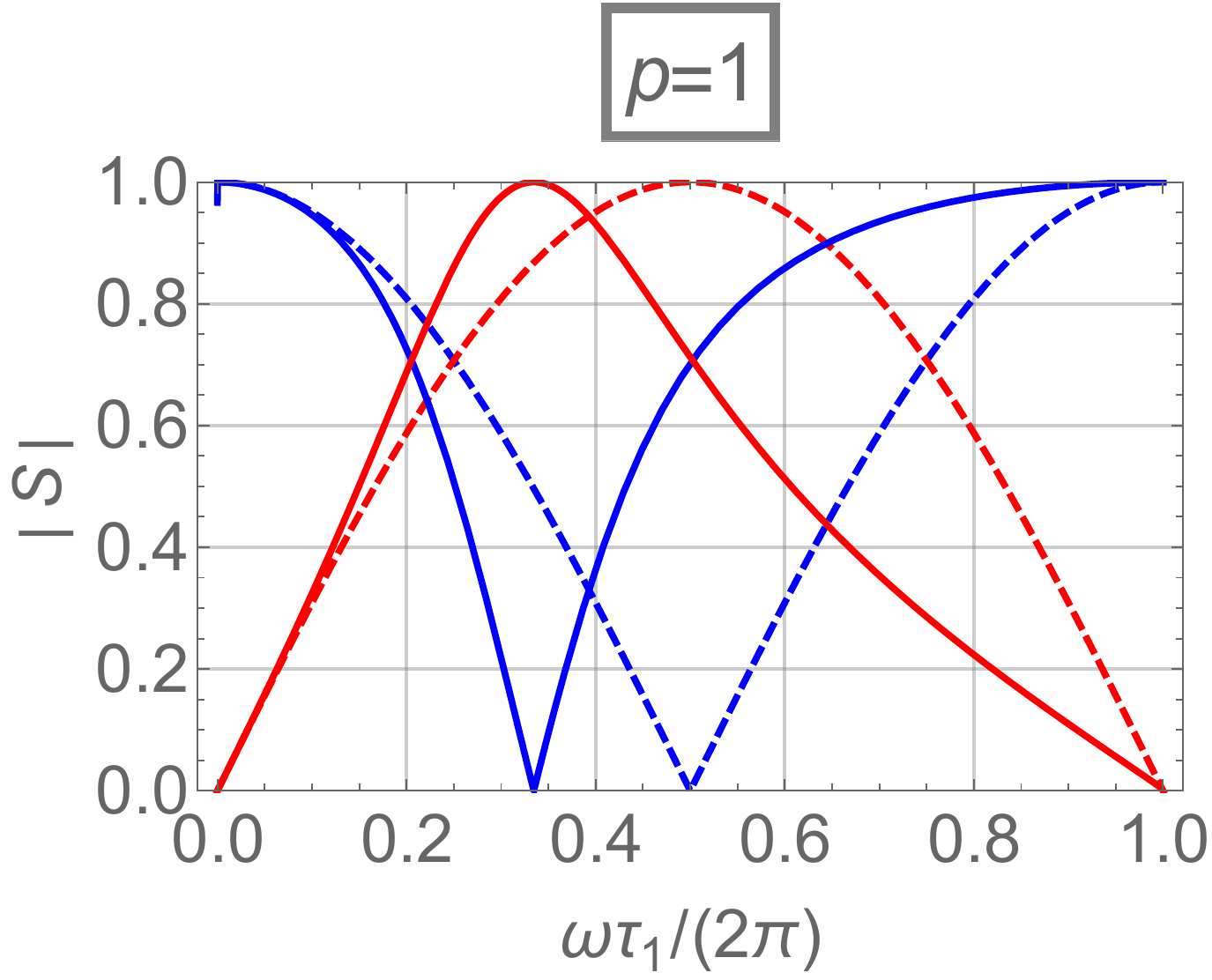}
\includegraphics[width=0.23\textwidth]{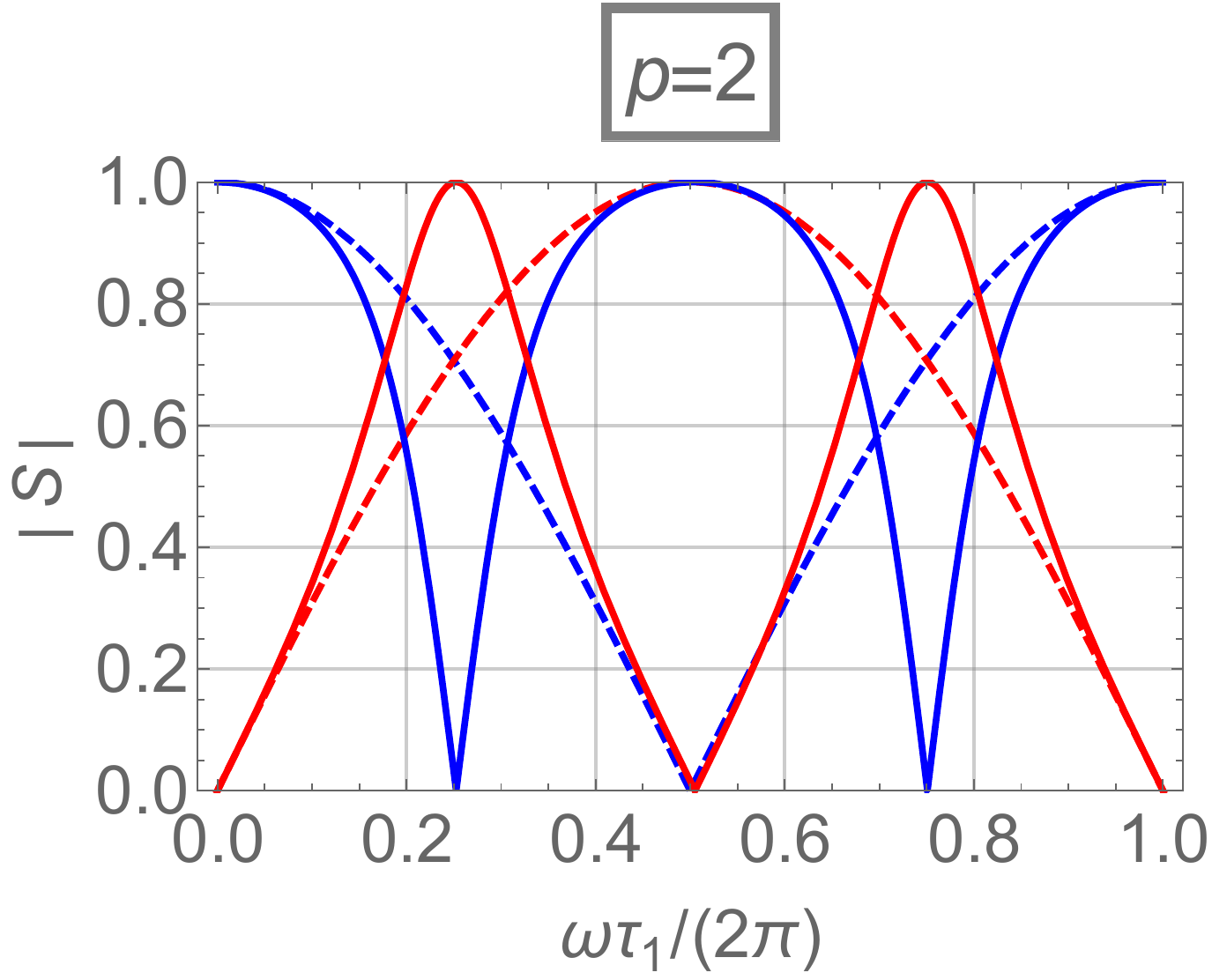}\\
\includegraphics[width=0.23\textwidth]{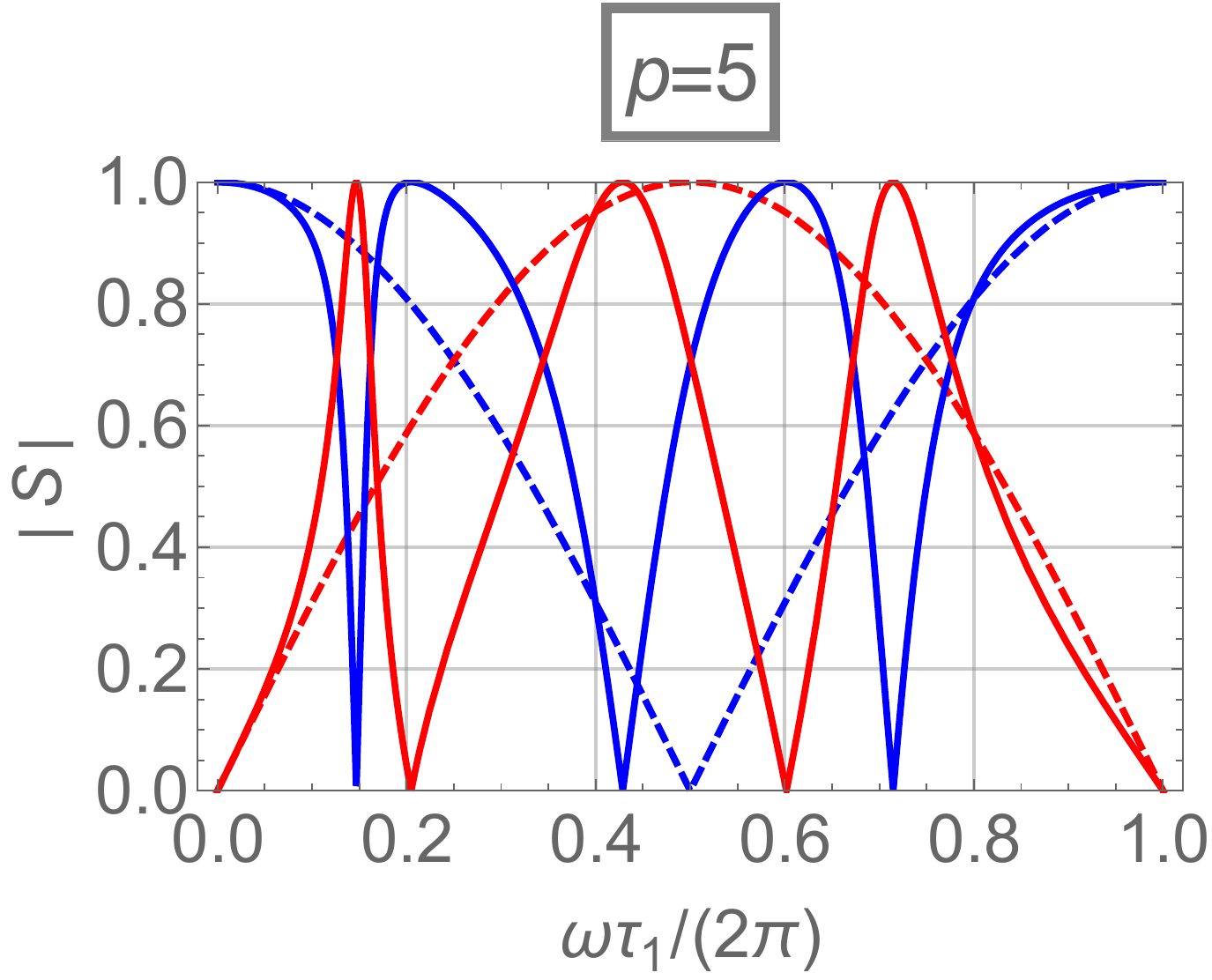}
\includegraphics[width=0.23\textwidth]{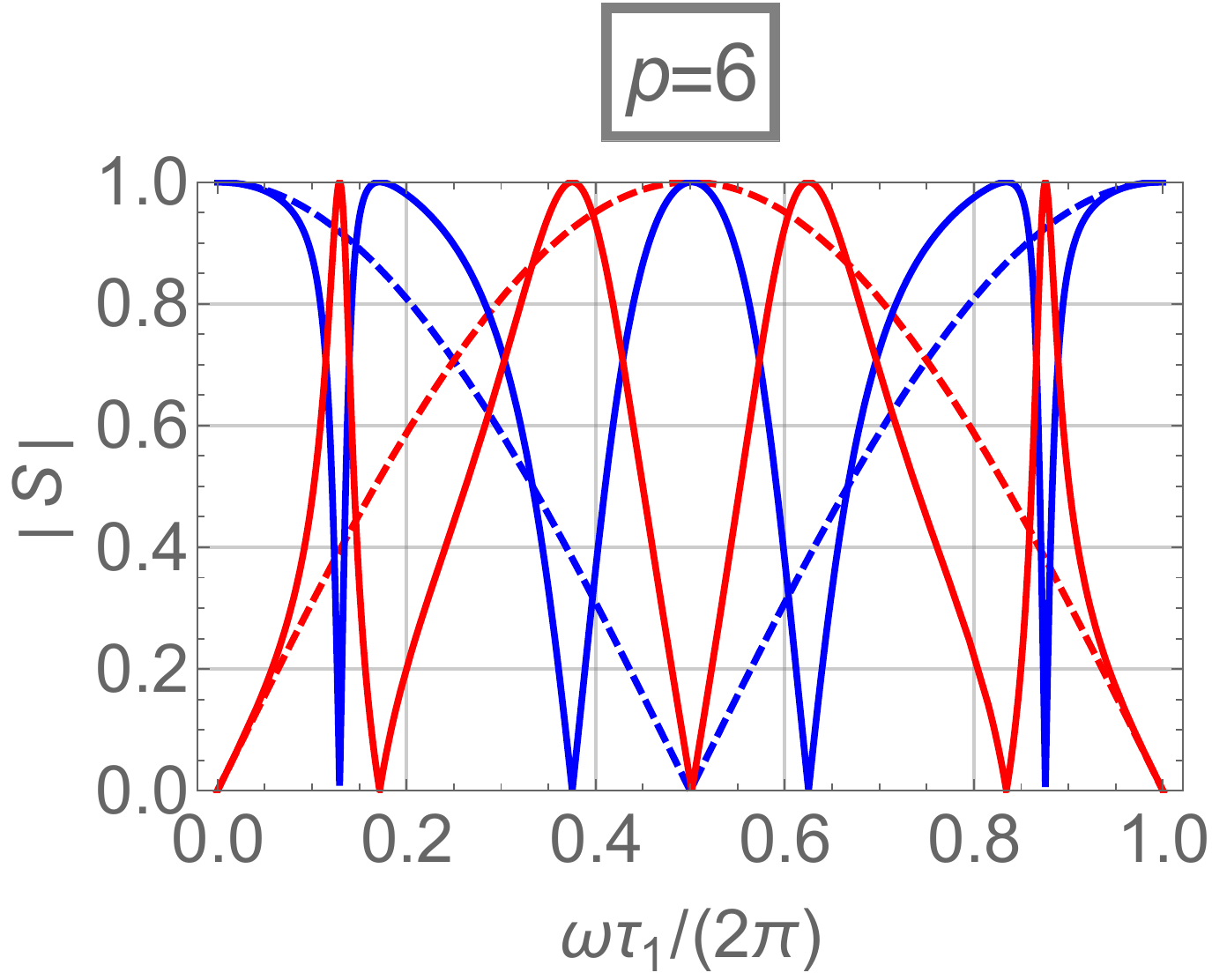}
\caption{Changes in the scattering parameters of a QH droplet due to a large value of the parasitic capacitor $C_{34}\equiv 2\sigma_{xy}T_1$. 
We used in the plots the same conventions used in Fig. \ref{fig:scattering_floating_par34}, but the solid lines are now obtained by using a higher value of the parasitic capacitor, $T_1/\tau_1=10$. 
%$p=2\tau_0/\tau_1$; Blue line $\left|S_{11}\right|$ red line $\left|S_{12}\right|$. The dashed line is ideal case of no parasitics and floating gates: envelope function obtained by setting $\tau_0=0$, We used $T_0=0$ $T_1/\tau_1=10$ .
}
\label{fig:scattering_floating_par34_10}
\end{figure}

% amplitudes of the additional Fano resonances gradually recover their original modulation, and the remaining resonances, associated with the limit $Z_{3,4}\rightarrow 0$, appear 

%the resonances $\tilde{\omega}_n$ are shifted back to $\omega_n$ and their bandwidth is restored to the original value in Eq.    (\ref{eq:bandwidth}); also, the amplitudes of the additional Fano resonances gradually recover their original modulation, and the remaining resonances, associated with the limit $Z_{3,4}\rightarrow 0$, appear.

%First of all, in this case, there are $\lceil p\rceil$ resonances: the remaining resonances are associated with the limit $Z_{3,4}\rightarrow 0$, and they appear only at finite $T_0$.
%For this reasons, a finite $T_1$ strongly degrades the performances of the floating configuration in Sec.\ref{sec:floating-gates} for long devices with $p\gg 1$, and in this limit the grounded configuration seems more reliable.
%In fact, as $T_0$ increases, the resonances $\tilde{\omega}_n$ are shifted back to $\omega_n$ and their bandwidth is restored to the original value in Eq.    (\ref{eq:bandwidth}); also, the amplitudes of the additional Fano resonances gradually recover their original modulation and the remaining resonances associated with the limit $Z_{3,4}\rightarrow 0$ appear.

It is also interesting to consider what happens when the ratio $T_1/\tau_1$ is large, and the two screening electrodes are shorted together and left floating.
% This makes no difference when $T_0\rightarrow\infty$, but it plays a role when the electrodes are left floating.
As discussed before, as $T_1/\tau_1$ increases, the resonances are shifted to lower frequencies, but the broadening of the central and of the Fano resonances respectively decreases and increases, until the peaks become comparable.
Then, depending on the parity of $p$, we arrive to different limiting situations, as shown in Fig. \ref{fig:scattering_floating_par34_10}. Interestingly, for even $p$, the device is perfectly reflecting at the original central resonance frequencies $\omega_n$.
This can be understood by realizing that at these frequencies all the electrodes act as short circuits, thus a voltage wave coming from one of the driving terminals travels back to the same terminal through the short connecting the two screening electrodes.
Also, note that the limit $T_1\rightarrow\infty$ can be used to model the effect of a back gate that covers the whole device, as discussed in Sec. \ref{sec:back-gate}.\\

We focus now on the capacitances that connect the driving electrodes to ground $C_{1G}, C_{2G}$. Again, we assume them to be equal and we introduce the corresponding charging time  $T_2 \equiv C_{1G}/(2\sigma_{xy})$.
These capacitors just degrade the performance of the device, regardless of how the screening electrodes are connected to ground.
In particular, when we increase $T_2$, the central resonances  are shifted to higher frequencies $\tilde{\omega}_n$, according to
\begin{equation}
\frac{(\tilde{\omega}_n-\omega_n) \tau_1}{4\pi(1+2n)}\approx\frac{T_2}{2\tau_1}+\left(\frac{T_2}{\tau_1}\right)^2,
\end{equation}
and at the same time their bandwidth decreases, until for high enough $T_2/\tau_1\gg 1$ all the current flows to ground and the device stops working.
The additional resonance peaks associated with the current flowing in the screening electrodes have qualitatively the same behavior.\\

\begin{figure}
a)\includegraphics[width=0.22\textwidth]{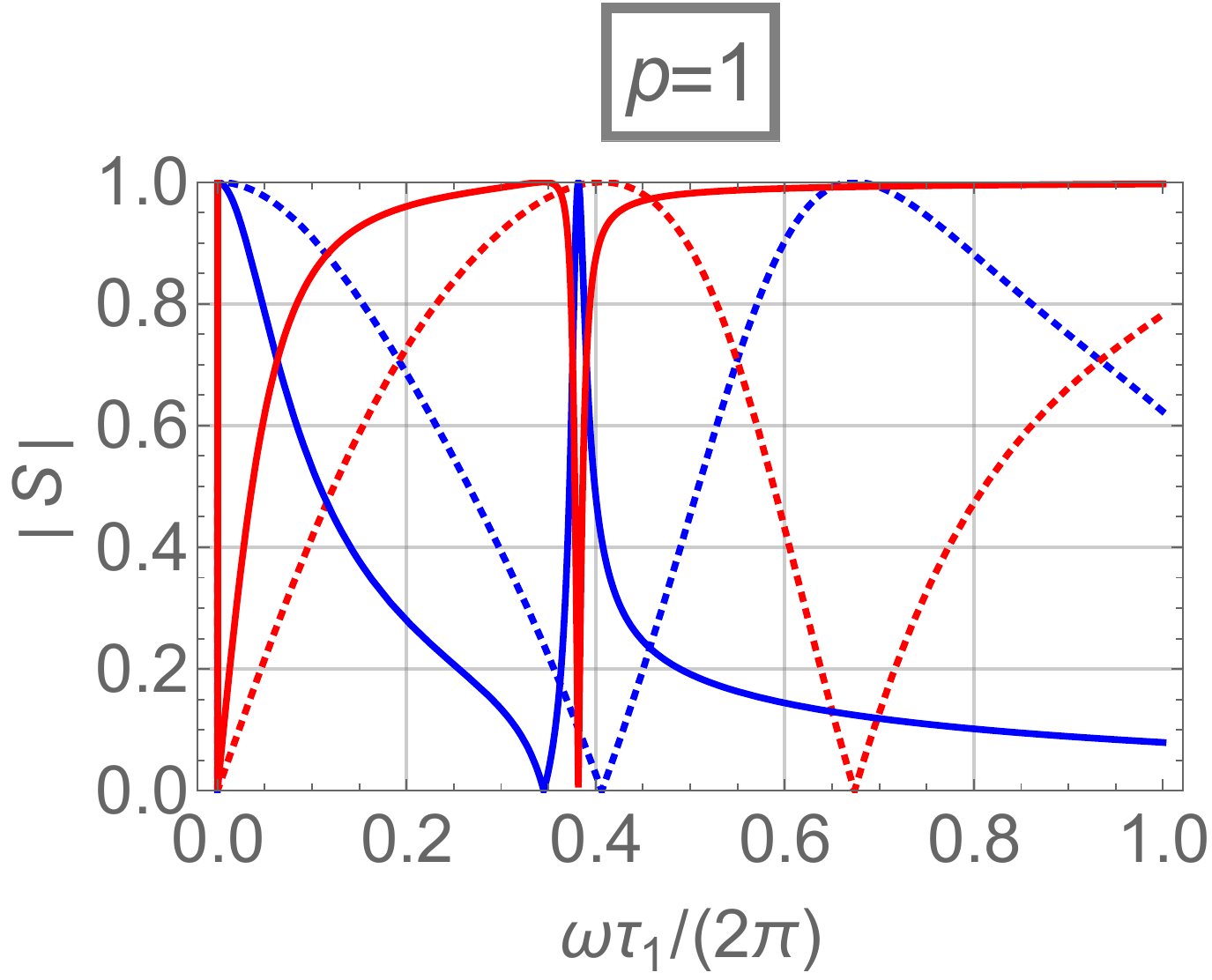}
b)\includegraphics[width=0.22\textwidth]{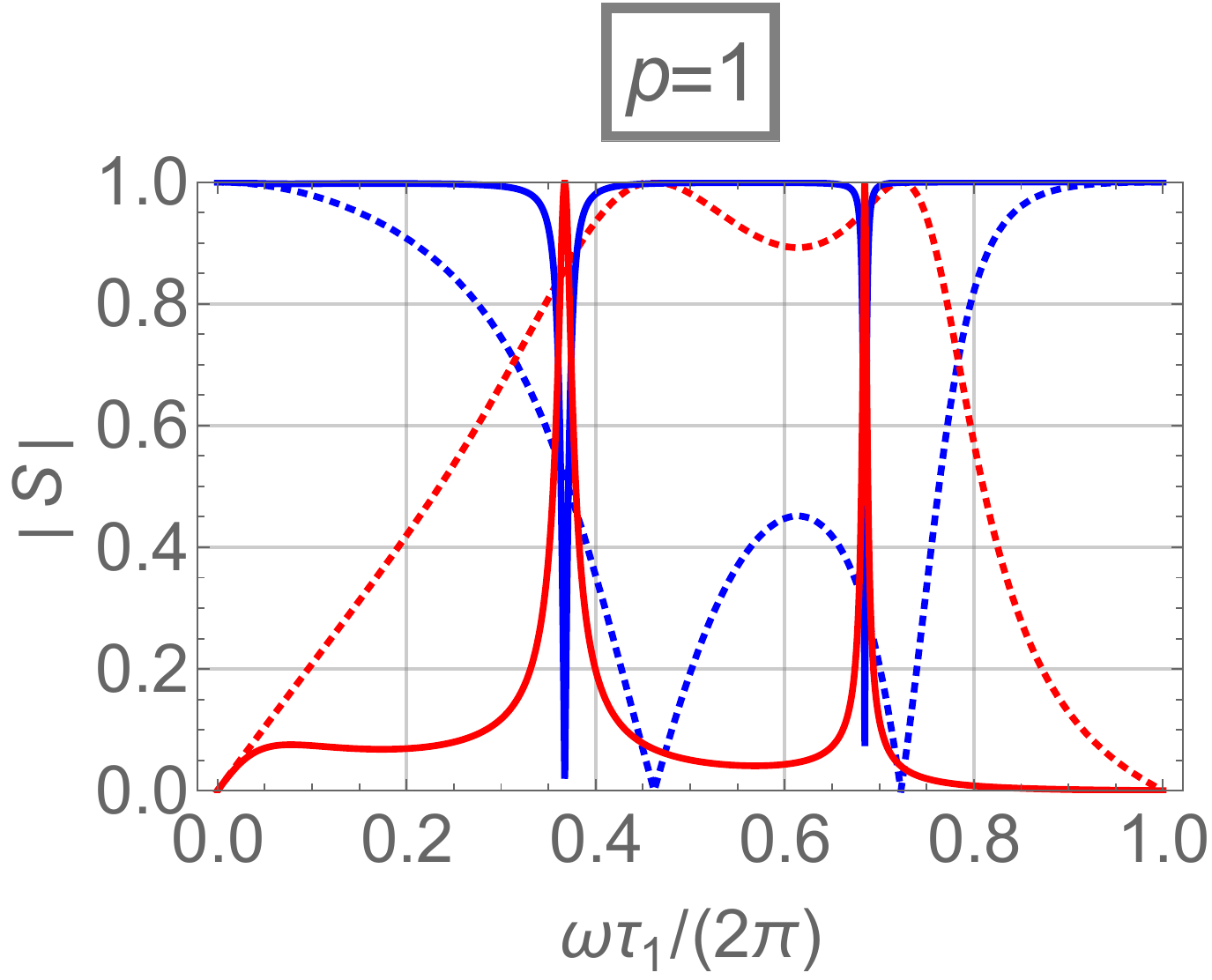}
\caption{Changes in the scattering parameters of a QH droplet due to the parasitic capacitors connecting adjacent electrodes $C_{13}=C_{14}=C_{24}=C_{24} \equiv 2\sigma_{xy}T_3$. 
In these plots, we assume $\alpha\equiv 2 \sigma_{xy}Z_0=1$ and a symmetric configuration  with $\tau_1=\tau_2$ and $\tau_3=\tau_4=\tau_1 p/2$. Here, we set $p=1$.
In a) and b) we consider respectively screening gates that are floating ($T_0\rightarrow 0$) and connected to ground ($T_0\rightarrow\infty$). 
The blue (red) solid line represents the absolute value of $\left|S_{11}\right|$ $(\left|S_{12}\right|)$ when $T_3/\tau_1=1$, while the dotted lines are obtained for $T_3/\tau_1=0.1$.
%$p=2\tau_0/\tau_1$; Blue line $\left|S_{11}\right|$ red line $\left|S_{12}\right|$. $p=1$ a) floating $T_0=0$ b) grounded $T_0\rightarrow\infty$. The dotted lines are with $T_3/\tau_1=0.1$ while the solid lines are for $T_3/\tau_1=1$.
}
\label{fig:scattering_floating_par13}
\end{figure}

Finally, we examine the effect of the capacitances connecting adjacent electrodes $C_{13},C_{14},C_{23},C_{24}$, and we take them to be equal and parametrized by the charging time $T_3\equiv C_{13}/(2\sigma_{xy})$.
If the screening electrodes are grounded, these capacitors simply connect the driving terminals to ground in the same way as $C_{1G}$ does, and so  $T_3$ and $T_2$ have the same effect on the response of the device.
In contrast, if the screening electrodes are floating, the response is affected in a way that resembles more the effect of $T_1$.
In particular, for small $T_3$, the central resonance peak is shifted to a lower frequency, approximately as in Eq.    ({\ref{eq:shift-resonance-frequency-34}}) with $T_1\rightarrow T_3$ and also the bandwidth shrinks in a similar way; additionally, Fano resonances begin to appear due to the finite amount of current flowing in the screening electrodes.
However, there is an important difference between the effect of $T_1$ and $T_3$: in this configuration, when $Z_1(\omega)\rightarrow\infty$ and the driving electrodes act as open circuits, a finite current can still flow to the opposite terminal through the parasitic capacitance.
Therefore, in contrast to what happens for $T_1$, the device is not perfectly reflecting at the frequencies $\omega_m$ in Eq.  (\ref{eq:cond_open}): as $T_3$ increases, these peaks are shifted to lower frequencies.
This consideration leads to a very different limit when $T_3/\tau_1$ is large.
In this situation, the device acts like a capacitor connecting the two driving terminals and it perfectly transmits the signal, except for residual resonances where the transmission drops and the reflection increases. Interestingly, when the value of $T_0$ is large, one obtains the opposite result and the device behaves as a perfect reflector, as shown in Fig. \ref{fig:scattering_floating_par13}.

%Figures

\subsection{Back gate \label{sec:back-gate}}

The transmission line shown in Fig. \ref{fig:transmission-line} has two distinct screening electrodes coupled to the edges of the QH material.
An alternative way of realizing a TL with similar features is to replace the two screening electrodes with a single back gate.
We comment here on how the circuit model in Fig. \ref{fig:circuit-model} is modified in this case.

Because there is only a single screening gate, the most apparent change with respect to before is that the screening terminals are always shorted together.
%This short can be modeled as a very high parasitic capacitor $C_{34}\propto T_1\rightarrow \infty$, whose effect is analyzed in Sec. \ref{sec:parasitics}; from the relative discussion, it appears more convenient to connect the back gate to ground, and thus we only focus on this case.
This short can be modeled as a very high value of the parasitic capacitor $C_{34}\propto T_1\rightarrow \infty$ shown in Fig. \ref{fig:circuit-parasitics}. From the discussion in Sec. \ref{sec:parasitics}, it appears more convenient to connect the back gate to ground, so that $C_{34}$ has no effect, and thus here we analyze only this configuration.

Additionally, we assume for simplicity that the contribution to the velocity due to the screening of the back gate $v_B$ is constant in the whole device, and so  $v_3=v_4=v_B$.
If the back gate extends below the whole QH material, the motion of the EMPs in the driving regions $1$ and $2$ is also affected by its presence.
The additional screening in these regions causes some interesting differences in the response;  in particular, the equation of motion (\ref{eq:motion-emp-local-velocity-app}) is modified in two ways.

First, the EMP velocities $v_{1,2}$ in Eq. (\ref{eq:EMP-vel-app}) are renormalized by $v_B$: the renormalized velocities $\tilde{v}_{1,2}$ are lower than $v_{1,2}$ and they can be estimated by \cite{HITLpt2}
\begin{equation}
\label{eq:renormal-valocity-BG}
\tilde{v}_{1,2}=\frac{1}{1/v_{1,2}+1/v_B}.
\end{equation}

Secondly, $v_B$ also modifies the inhomogeneous driving term of the equation of motion, that now reads
\begin{equation}
\label{eq:renormal-voltage-BG}
\sigma_{xy} \partial_y V_a(y,\omega)\rightarrow \sigma_{xy}\partial_y\left(\left( 1- \frac{\tilde{v}(y)}{v_B}\right)  V_a(y,\omega)\right)
\end{equation}

In a similar way, the amount of current that flows in the driving electrodes in Eq.    (\ref{eq:current-ith-electrode}) is also modified by the presence of the back gate, and it acquires the same multiplicative prefactor $\left( 1-\tilde{v}_{1,2}/v_B\right)$ as the driving voltage.

%A more detailed justification for these modifications is postponed to Sec. \ref{sec:EMP-gated}.
A detailed justification for these modifications and a correction of these equations valid in the sharp edge limit can be found in \cite{HITLpt2}.
Here, we just point out that they are consistent with the local capacitance model of \cite{Viola-DiVincenzo}.
In this model, the EMP velocity is inversely proportional to a local capacitance (per unit length) function $c(y)$ which quantifies the Coulomb coupling between the EMPs and the electrodes. In particular, for smooth QH edges \cite{JohnsonVignale,Volkov}, one can roughly estimate the coupling to the driving electrodes 1,2 and to the back gate as $c_{1,2,B}\propto \epsilon_S/d_{1,2,B}$, where $\epsilon_S$ is the dielectric constant of the medium and $d_i$ is the distance of the $i$th metal plate from the boundary of the QH material. 
The renormalization of the velocities is then straightforwardly understood by considering that the total effective capacitance from the QH edge to ground is composed of the two capacitances $c_{1,2}$ and $c_B$ in parallel, leading to $\tilde{v}_{1,2}\propto 1/ (c_{1,2}+c_B)$. Considering that $v_{1,2,B}\propto 1/c_{1,2,B}$, one can obtain Eq. (\ref{eq:renormal-valocity-BG}).
Also, the presence of two capacitors in the driving regions immediately explains the partition of voltages and currents: only a factor $c_B^{-1}/(c_{1,2}^{-1}+c_B^{-1})=1-\tilde{v}_{1,2}/v_B$ of the total applied voltage and of the total current is relevant for the response.

%With this interpretation, the renormalization of the velocities is caused by t  to the total parallel capacitance per unit length $c$ coupling the terminal and the back electrode, i.e. $\tilde{v}_{1,2}\propto 1/ (c_{1,2}+c_B)$.
%Also, in this model, the presence of two capacitors in the terminal regions immediately leads to a partition of voltage and current: only a factor $c_B^{-1}/(c_{1,2}^{-1}+c_B^{-1})=1-\tilde{v}_{1,2}/v_B$ of the total applied voltage and of the total resulting current is relevant for the response.

The $(i,j)$ matrix element of the port admittance in this configuration can be easily derived from the same matrix element of the port admittance obtained when the two screening electrodes are grounded (i.e. the $2\mathrm{x}2$ upper block of the terminal admittance (\ref{eq:admittance-4x4})) by the following substitutions
% upper $2\times 2$ block of the terminal admittance matrix $Y_{ij}$ in Eq.    (\ref{eq:admittance-4x4}) obtained when the two screening is grounded then modified by the two substitutions
\begin{subequations}
\label{eq:substitution-bg}
\begin{flalign}
\label{eq:renormalized-timescale}
\tau_{1,2} &\rightarrow \tilde{\tau}_{1,2}\equiv \frac{L_{1,2}}{\tilde{v}_{1,2}},\\
\sigma_{xy} & \rightarrow \sigma_{xy}\left(1-\frac{\tilde{v}_{i}}{v_B}\right)\left(1-\frac{\tilde{v}_{j}}{v_B}\right).
\end{flalign}
\end{subequations}

Note that these modifications are qualitatively different from the parasitic capacitance $C_{1G}$ to ground described in Sec. \ref{sec:parasitics}, which simply adds the term $i\omega C_{1G}$ to the diagonal elements of the port admittance matrix.

To understand the effect of these substitutions, we focus on a symmetric configuration and we specialize to $\tilde{v}_{1}=\tilde{v}_{2}$.
In this case, the response is qualitatively similar to the one described in Sec. \ref{sec:grounded}, but with different propagation times $\tilde{\tau}_{1,2}$,  and, most importantly, with an increased characteristic impedance
\begin{equation}
\frac{1}{2\sigma_{xy}}\rightarrow \frac{1}{2\sigma_{xy}}\left(1-\frac{\tilde{v}_{1}}{v_B}\right)^{-2},
\end{equation}
due to the partition of current and voltage between the driving electrodes and the back gate.

\subsection{Additional modes and Coulomb drag\label{sec:additional_modes_coulomb_drag}}

To conclude the analysis of a single QH droplet, we comment here on how the performance of the device is influenced by additional plasmonic modes and Coulomb interactions between different edges; for simplicity we analyze only a symmetric configuration.

A detailed analysis of the influence of slower plasmonic modes in non-reciprocal QH devices can be found in \cite{QEMP}.
For TLs, the results are qualitatively similar: the equation of motion (\ref{eq:motion-emp-local-velocity-app}) is replaced by a set of $N_{\text{modes}}$ independent equations, where $N_{\text{modes}}$ is the number of modes.
These equations of motion are characterized by different plasmon velocities $v_n(y)$ and couplings to external electrodes $\sigma_{xy}\rightarrow \sigma_{xy} \Gamma_n$; the coupling constant $ 0\leq\Gamma_n\leq 1$ is higher for faster modes and its precise value depends strongly on the electrostatic model of the edge.
%The main effect is to add Fano resonances to the response of a single mode, with rescaled velocity and coupling. This comes from the fact that the equation of motion (\ref{eq:motion-emp-local-velocity-app}) is substituted by a set of $N$ independent equations, characterized by different velocities $v(y)_n$ and coupling to external electrodes $\sigma_{xy}\rightarrow \sigma_{xy} \Gamma_n$, where the coupling function $ 0\leq\Gamma_n\leq 1$ is higher for faster modes and its precise value depends strongly on the intra edge interactions.
%Also, the EMPs acquire a more complicated charge density because of interactions, and this has to be accounted in the integral in Eq.    (\ref{eq:current-ith-electrode}).
Also, because of interactions, the EMPs acquire a more complicated charge density structure, and this has to be accounted for when computing the current from the integral in Eq. (\ref{eq:current-ith-electrode}).

These modifications can be modeled by an equivalent circuit composed of $N_{\text{modes}}$ copies of the circuit in Fig. \ref{fig:circuit-model} connected in parallel. These copies describe the response of different plasmonic modes and, consequently, they are characterized by different propagation times and characteristic impedances.
The presence of additional parallel circuits with different resonance frequencies leads to Fano-like resonances in the $S$-parameters of the device \cite{QEMP}.
The effect of slower EMP modes is negligible for low filling factors, that maximize the characteristic impedance, when the edge of the QH droplet is abruptly defined. \\

The Coulomb interactions between EMPs localized at opposite edges modify the equation of motion of the excess charge density in a similar way.
When a positive charge wave is launched at one edge, because of the inter-edge interactions, it drags with it a small amount of negative charge at the opposite edge  \cite{Circuit_EMP,HITLpt2}; because of this Coulomb drag, the EMP propagation velocity is lowered. 
The response of the system can be described by considering two EMP modes with charge densities with opposite sign that propagate in opposite direction: these modes produce currents flowing in the same direction.
A microscopic analysis of this situation is presented in \cite{HITLpt2}; here we limit ourselves to a discussion of the effect at the circuit level.

In analogy to before, the equivalent circuit model of the device is made up of two circuits as the one in Fig. \ref{fig:circuit-model}  connected in parallel.
If we consider a symmetric setup, where the two screening electrodes 3,4 have the same length and the velocities $\tilde{v}_{3,4}$ (renormalized by the inter-edge interactions) are equal, the propagation times in the two circuits are also equal.
The frequency dependent part of the admittance matrix can then be factorized and consequently we are effectively left with a single circuit with characteristic impedance $1/(2\sigma_{xy})$. For this reason, at the circuit level, the response of symmetric setups is not qualitatively altered by the Coulomb drag and its only effect is to decrease the value of the EMP velocity.
%Differently from before, however, the absolute value of the propagation velocity of the two modes can have the same value.
%For example, let us consider the nanowire in Fig.\ref{fig:transmission-line} with side electrodes placed symmetrically, such that $v_3=v_4$.
%In this case, the time-scales $\tau_i$ are equal for the two modes, and the frequency dependence of admittances of the two circuits can be factored out, leaving a single circuit with renormalized velocity and characteristic impedance $1/(2\sigma_{xy})$.
% From a mathematical point of view it is easy to understand this result by considering that the admittance in Eq.    (\ref{eq:admittance-4x4}) satisfies $Y_{ij}^{\circlearrowleft}(\omega)=-Y_{ij}^{\circlearrowright}(-\omega)$, where the arrow indicates the direction of the circulator in Fig.\ref{fig:circuit-model}; in other words, the additional EMP mode, with opposite charge, velocity and direction of propagation gives up to a constant the same admittance as before.

\section{Lumped element transmission line\label{sec:meta-material-tl}}

\begin{figure}
a)\includegraphics[width=0.45\textwidth]{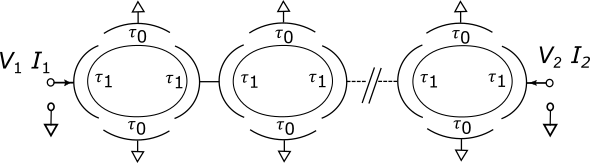}\\
b)\includegraphics[width=0.45\textwidth]{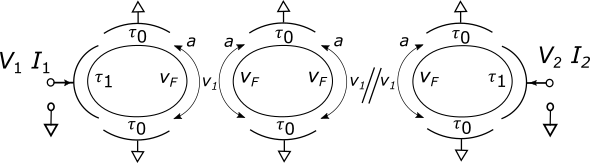}
\caption{Meta-material transmission lines.
The meta-materials are composed of a chain of identical QH droplets coupled in different ways. 
The electrostatic coupling to the different metal electrodes is quantified by the corresponding propagation time $\tau_i$.
Here, we focus on symmetric droplets characterized by only two propagation times $\tau_{0,1}$ and whose screening electrodes are connected to ground. 
In a) the coupling between adjacent droplets is mediated by a metal electrode of negligible length, while in b) the droplets are coupled via unscreened Coulomb interactions. In the latter case, the coupling region is characterized by its length $a$ and by the intra- and inter-edge velocities, $v_F$ and  $v_1$, respectively.}
\label{fig:coupling_schemes_droplets}
\end{figure}

%\begin{figure}
%\centering
%\begin{subfigure}[a]{0.4\textwidth}
%\includegraphics[width=0.45\textwidth]{coupling_schemes_droplets_a}
%\end{subfigure}
%\begin{subfigure}[b]{0.4\textwidth}
%\includegraphics[width=0.45\textwidth]{coupling_schemes_droplets_b}
%\end{subfigure}
%\caption{Meta-material transmission lines.
%The meta-materials are composed of a chain of identical QH droplets coupled in different ways. 
%The electrostatic coupling to the different metal electrodes is quantified by the corresponding propagation time $\tau_i$.
%Here, we focus on symmetric droplets characterized by only two propagation times $\tau_{0,1}$ and whose screening electrodes are connected to ground. 
%In a) the coupling between adjacent droplets is mediated by a metal electrode of negligible length, while in b) the droplets are coupled via unscreened Coulomb interactions. In the latter case, the coupling region is characterized by its length $a$ and by the intra and inter-edge velocities, $v_F$ and  $v_1$, respectively.}
%\label{fig:coupling_schemes_droplets}
%\end{figure}

In this section, we discuss another possible implementation of a TL.
In particular, we study the microwave response of a meta-material composed of a chain of $N$ QH droplets.
For simplicity, we restrict our analysis to droplets whose screening electrodes are connected to ground, and we assume that each droplet interacts only with the adjacent ones.
Also, we neglect the possible parasitic coupling between electrodes and ground.
The way in which adjacent droplets are coupled with each other strongly influences the behavior of the whole system.
Here, we focus on the two configurations shown in Fig. \ref{fig:coupling_schemes_droplets}.

We begin by considering the setup in  Fig.  \ref{fig:coupling_schemes_droplets} a), where there are $N$ identical QH droplets coupled via a thin metal electrode.
% In this case, the coupling is mediated by a thin metal electrode. a cascade of the QH devices described in Sec. \ref{sec:distributed-tl} that is a coupling mediated by a thin metal electrode.
% We neglect the possible parasitic coupling of this metal to ground, which only gives a loss of the transmitted signal, as described in Sec. \ref{sec:parasitics}. 
We restrict to a symmetric configuration, with $\tau_1=\tau_2$ and $\tau_3=\tau_4\equiv\tau_0$, see Fig. \ref{fig:circuit-model}, so that each of the droplets acts as the unconventional TL in Fig. \ref{fig:circuit-grounded}.
If the metal lead connecting two adjacent droplets is short enough, we can also neglect the phase accumulated by the signal in passing from one droplet to the next one, and the total transfer matrix of the chain $\mathcal{T}^N$ is obtained simply by taking the $N^{\mathrm{th}}$ power of the transfer matrix $\mathcal{T}$ in Eq.    (\ref{eq:ABCD_matrix_grounded}), which models a single droplet.

To characterize the response of the device, we compute the scattering parameters using $\mathcal{T}^N$.
%For simplicity, we consider the system to be symmetric and thus we set for all droplets 
%\begin{equation}
%Z_1=Z_2=-\frac{i}{2\sigma_{xy}}\cot\left(\frac{\omega \tau_1}{2}\right),
%\end{equation}
In doing so, we assume for simplicity that the impedance of the chain is matched to the external microwave circuitry, i.e. $Z_0=1/(2\sigma_{xy})$.
Note that to compensate for the typical high mismatch between conventional microwave circuits ($Z_0\approx 50\mathrm{\Omega}$) and the QH material, one can modify the first and last droplets to match the two impedances at the specific frequency of operation, as described in Sec.  \ref{sec:grounded}. 
When the impedance mismatch is high, such a construction is expected to have a low bandwidth. The bandwidth can be improved by using a tapered construction \cite{Pozar}, for example, a cascade of $M$ droplets, the $m^{\mathrm{th}}$ of which matches the input impedance $Z^{\text{in}}_m=Z_0+m/(2\sigma_{xy}M)$ to the closer output impedance $Z^{\text{out}}_{m}=Z_0+(m+1)/(2\sigma_{xy}M)=Z^{\text{in}}_{m+1}$.\\

\begin{figure}
\includegraphics[width=0.36\textwidth]{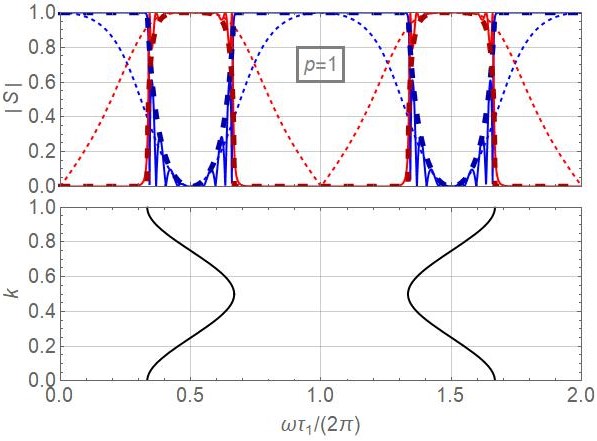}
\includegraphics[width=0.36\textwidth]{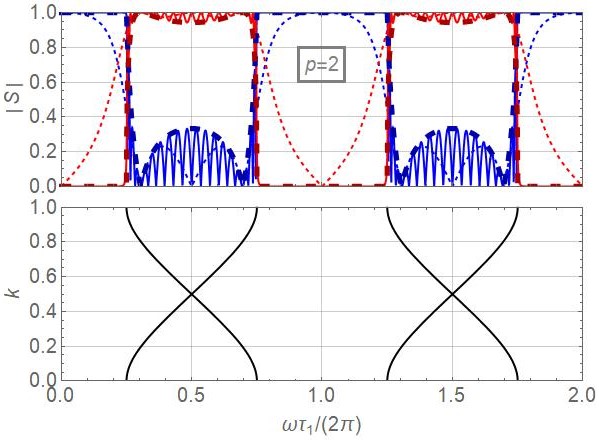}\\
\includegraphics[width=0.36\textwidth]{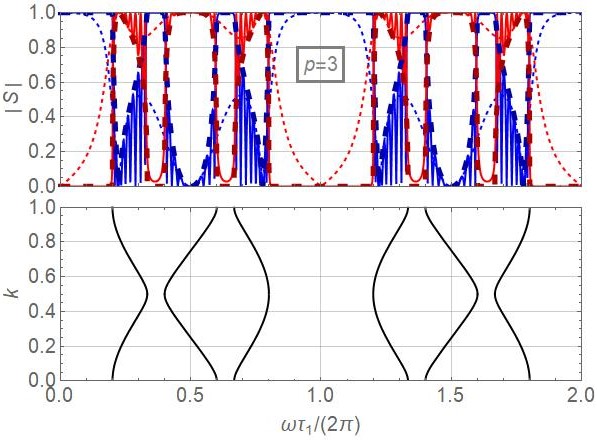}
\includegraphics[width=0.36\textwidth]{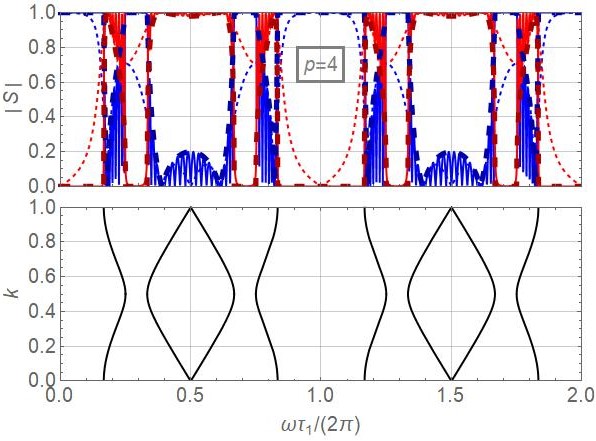}
\caption{
Absolute value of the $S$ parameters and dispersion relation for the meta-material transmission line in Fig. \ref{fig:coupling_schemes_droplets} a).
The different plots are obtained with different values of $p=2\tau_0/\tau_1$ and assuming a matched circuit $\alpha=1$.
The blue (red) solid lines in the top sub-plots are the reflection (transmission) parameters $\left|S_{11}\right|$ $(\left|S_{12}\right|)$ obtained for a chain comprising $N=10$ unit cells.
The thin dashed lines represent the response for a single droplet and the thick dashed lines in darker colors are the smooth envelope functions modulating the fast resonances.
In the sub-plots at the bottom, we show the band dispersion $k(\omega)$ (Eq. (\ref{eq:band-structure})) of the corresponding periodic chain.}
\label{fig:scattering_band_chain}
\end{figure}

The absolute value of the scattering parameters for different values of the ratio of propagation times $p$, defined in Eq. (\ref{eq:p-def}), are shown in Fig.  \ref{fig:scattering_band_chain}.
Similar to the situation for a single droplet, the response can be decomposed into fast oscillations modulated by a smooth envelope function.
However, as $N$ increases, the $S$ parameters begin to display new interesting features. 
In particular, for certain frequencies, the transmission is not allowed and the reflection is maximal: these frequency gaps can be understood in terms of band structure.

When $N$ identical blocks characterized by a transfer matrix $\mathcal{T}$ are cascaded, the total transfer matrix $\mathcal{T}^N$ can generally be  written as
\begin{equation}
\mathcal{T}^N=\mathcal{M}  
	\begin{pmatrix}
        \lambda_+^N 	& 0 		\\
        0 			& \lambda_-^N 
    \end{pmatrix}
    \mathcal{M}^{-1},
\end{equation}
where $\mathcal{M}$ and $\lambda_{\pm}=\mathcal{T}_{11}\pm\sqrt{\mathcal{T}_{11}^2-1}$ are respectively the matrix of column eigenvectors and the eigenvalues of $\mathcal{T}$.
To find the band structure of the QH chain, we apply periodic boundary conditions, leading to the condition on the eigenvalues
%If we apply periodic boundary condition to this system, we get the condition on the eigenvalues
\begin{equation}
\label{eq:lambda-k}
\lambda_{\pm}=e^{i 2\pi n/N},
\end{equation}
with $n$ being an integer number between $0$ and $N-1$. 
%For the chain of QH droplets, the eigenvalues $\lambda_{\pm}$ depend on frequency, and thus this condition 
%For an infinite chain, we can introduce a real parameter $k\in [0,1)$ that physically plays the role of a crystal momentum and that quantifies the phase that a voltage wave accumulates from one droplet to the next one, in analogy to the Bloch-quasi momentum in electronic band theory.
For an infinite chain, we can introduce a real parameter $k\in [0,1)$ which quantifies the phase accumulated by a voltage wave in one unit cell, in analogy to the Bloch-quasi momentum in electronic band theory. In this context, the matrix $\mathcal{T}$ is sometimes referred to as the Floquet matrix.
From Eqs. (\ref{eq:ABCD_matrix_grounded}) and (\ref{eq:lambda-k}), it is straightforward to find the dispersion relation $\omega(k)$
\begin{equation}
\label{eq:band-structure}
\cos(2\pi k)=\mathcal{T}_{11}=\frac{\sin \left((1+p)\frac{\omega \tau _1 }{2} \right)}{\sin \left(\frac{\omega \tau _1 }{2}\right)}.
\end{equation}

In Fig. \ref{fig:scattering_band_chain}, we plotted  the dispersion for several values of $p$.
There are in general $\lceil p\rceil$ bands for each frequency period $\omega\tau_1/(2\pi)\in[n,n+1)$.
Comparing the $S$ parameters and the band structure, one can immediately verify that the transmission drops quickly to zero in the bandgaps and at these frequencies the device becomes perfectly reflecting.

More generally, it is well-known from solid state theory that the transmission probabilities in open crystals, with a number of unit cells $N\gg 1$, can be understood in terms of the band dispersion and the transmission of a single unit cell, see e.g. \cite{Barra}.
Analogously, in our case, the ratio between the power transmitted to the end port and power applied to the initial one, i.e. $\left|S_{12}^N\right|^2$ can be written in the revealing form
\begin{equation}
\label{eq:total-power-transmitted-N}
\left|S_{12}^N\right|^2 =\left(1+\frac{\left|S_{11}\right|^2}{\left|S_{12}\right|^2}\frac{\sin(2\pi k N)^2}{\sin(2\pi k)^2}\right)^{-1},
\end{equation}
where the crystal momentum $k$ is here a continuous function of frequency defined from the band expression (\ref{eq:band-structure}) and $S_{ij}$ are the $S$ parameters of a single QH droplet (Eq. (\ref{eq:s-par-ground})).
Also, since we assume a lossless network, the total reflected power is immediately obtained by $\left|S_{11}^N\right|^2+\left|S_{12}^N\right|^2=1$.
$\left|S_{11}^N\right|$ and $\left|S_{12}^N\right|$ are shown in Fig. \ref{fig:scattering_band_chain}.

From Eq.    (\ref{eq:total-power-transmitted-N}), it follows immediately that the resonances in the total transmission (and reflection) occur when the single droplet is perfectly reflecting, i.e. $\left|S_{11}\right|=0$, and when the crystal momentum takes the fractional values $k=n/N$; this latter condition corresponds to the eigenfrequencies of the periodic chain.

Additionally, the smooth envelope function modulating the amplitude of the resonances, can be obtained in the allowed frequency range by setting $\sin(2\pi k N)= 1$.
Note that this envelope has a different behavior depending on the parity of $p$ at the central resonances $\omega_n$ defined in Eq. (\ref{eq:cond_short}). 
This difference can be understood mathematically by observing that in the vicinity of $\omega_n$, the Bloch contribution reduces to $\left|\sin(2\pi k)\right|\rightarrow \left| \sin(\pi p/2)\right|$ and $\left|S_{11}\right|$ has the form discussed in Sec. \ref{sec:grounded} and shown in Fig. \ref{fig:scattering_grounded}. 

In particular, when $p$ is an integer odd number, the momentum contribution is maximal (and equal to one), while $\left|S_{11}\right|$ increases quadratically from zero as a function of $\omega$.
Therefore, close to the central resonance peaks, $\left|S_{11}\right|$ and the envelope function of $\left|S_{11}^N\right|$ have the same quadratic frequency dependence. 
In contrast, when $p$ is an integer even number, both $\left|S_{11}\right|$ and the momentum contribution vanish linearly as $\omega$ approaches $\omega_n$. For this reason, the modulating function of $\left|S_{11}^N\right|$ has a finite limit at $\omega\rightarrow\omega_n$, leading to a finite value of the reflection coefficient of the meta-material TL.
Physically, this reflection is related to the finite back-scattering that can occur at $\omega_n$ because of the crossing of bands with opposite curvature.

This band crossing is a physically interesting phenomenon, but from a technological point of view, the advantage of the meta-material  over the single droplet can be appreciated by considering $p=1$.
This case is particularly convenient for achieving an high bandwidth, since at $\omega_n$ there is a sweet spot where the reflection increases quadratically in $\omega$, specifically as 
\begin{equation}
\left|S_{11}^N\right|\approx\left|S_{11}\right|\approx\frac{3}{8}\left(\omega -\omega_n\right)^2\tau_1^2.
\end{equation}

If we assume the EMP velocity to be constant along the whole perimeter, to obtain $p=1$, we need the driving electrodes to be two times longer than the screening electrodes.
Then, a TL composed of a single droplet has a quite inconvenient aspect ratio when $p=1$, and to manufacture longer devices we require a higher value of $p$; in this case, however, the bandwidth varies only linearly in $\omega$, see Eq.    (\ref{eq:bandwidth}).
In contrast, in the meta-material TL, the effective propagation length can be tuned arbitrarily by varying the number of unit cells $N$, and so, using as unit cell a QH droplet with $p=1$, it is possible to implement a long TL while preserving the quadratic frequency dependence of the bandwidth.

\begin{figure}
\includegraphics[width=0.4\textwidth]{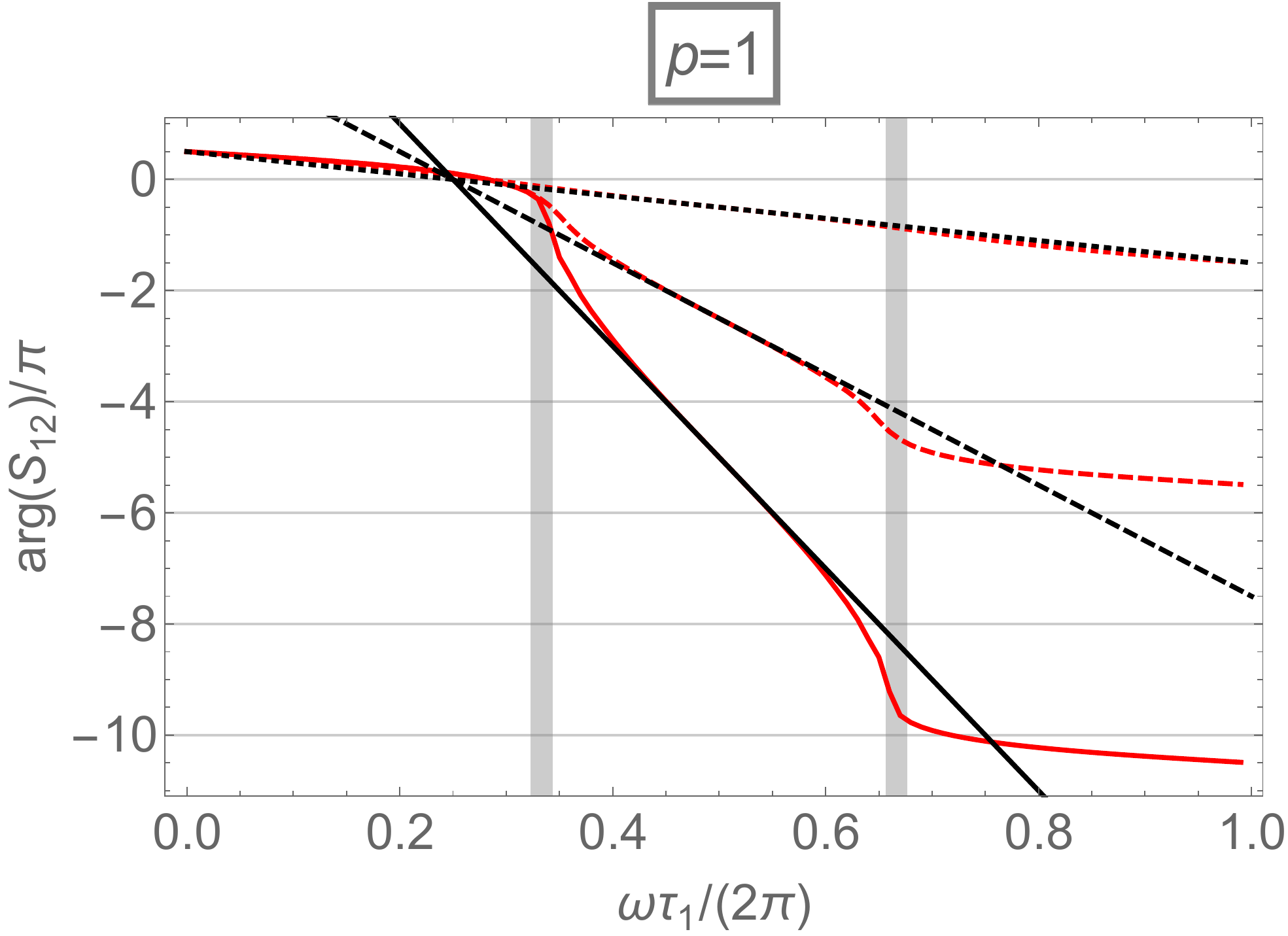}
\caption{\label{fig:phase-metamat} Phase of the transmission coefficient $S_{12}$ of the meta-material in Fig. \ref{fig:coupling_schemes_droplets} a). The red lines are the  phases of $S_{12}$, the black lines are obtained by using  the  linear approximation in Eq. (\ref{eq:phase-ground-meta-p1}). In the plot, we used a different number $N$ of unit cells, in particular $N=1,5,10$ for dotted, dashed and solid lines, respectively. The thick vertical lines at $\omega\tau_1=2\pi/3$ and $\omega\tau_1=4\pi/3$ represent the edges of the transmission band.}
\end{figure} 

For this configuration, we also examine the phase of the transmission coefficient $S_{12}^N$; the results are shown in Fig. \ref{fig:phase-metamat}.
Close to $\omega_n$, the phase $\arg(S_{12}^N)$ varies linearly in frequency, and it can be approximated as
\begin{equation}
\label{eq:phase-ground-meta-p1}
\arg(S_{12}^N)\approx N\frac{\pi}{2}-N\omega\tau_1.
\end{equation}
This dependence can be understood by comparing this result to Eq. (\ref{eq:phase-ground}): the phase $\arg(S_{12}^N)$ accumulated in passing through a system of $N$ cascaded QH droplets is simply $N$ times the phase $\arg(S_{12})$ accumulated in a single unit cell.
The linear frequency dependence of $\arg({S_{12}^N})$ guarantees that the device is not dispersive.\\

\begin{figure}
\includegraphics[scale=0.22]{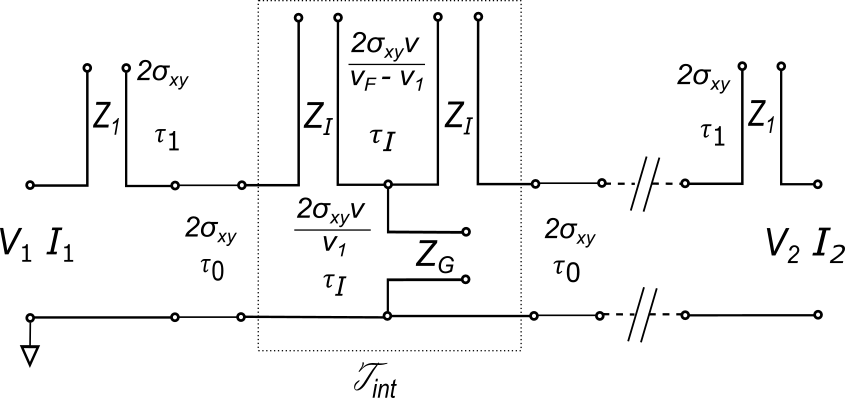}
\caption{Equivalent circuit model of the meta-material shown in Fig. \ref{fig:coupling_schemes_droplets} b).
The Coulomb interactions between adjacent droplets are modeled by the transfer matrix $\mathcal{T}_{\text{int}}$ in Eq.    (\ref{eq:transfer-int-coulomb-drag}). This is represented by a $T$-junction composed of the delay lines $Z_{I}$ and $Z_{G}$, characterized by the propagation time $\tau_I\equiv a/v$ and by  the characteristic impedances $(v_F-v_1)/(2\sigma_{xy} v)$ and  $v_1/(2\sigma_{xy} v)$, respectively. Here, $v=\sqrt{v_F^2-v_1^2}$.
This $T$-junction is repeated $N-1$ times, where $N$ is the number of QH droplets. Each junction is connected to the next one by an ideal transmission line with characteristic impedance $1/(2\sigma_{xy})$ and propagation time $\tau_0$. The capacitive coupling to the external electrodes is assumed to be the  same for the initial and final droplets and is modeled by the delay lines $Z_{1}$ (Eq. (\ref{eq:terminal-impedance-frequency})) with a propagation time $\tau_1$ and a characteristic impedance $1/(2\sigma_{xy})$.
}
\label{fig:circuit_model_coulomb_drag}
\end{figure}

We can also examine different coupling schemes between adjacent droplets.
In particular, we focus now on droplets coupled via Coulomb interactions between edge states, as shown in Fig. \ref{fig:coupling_schemes_droplets} b).
The length $a$ of the coupling regions is assumed to be much longer than the distance between adjacent droplets.
In these regions, there is a Coulomb drag between the EMPs localized at the edges of the two droplets; this drag is determined by the interplay of intra- and inter-edge Coulomb interactions, that are parametrized respectively by the velocities $v_F$ and $v_1$.
In particular, as discussed in Sec. \ref{sec:additional_modes_coulomb_drag} and in Ref. \cite{HITLpt2}, a finite $v_1$ leads to the presence of counterpropagating modes with opposite charge, moving with a renormalized velocity $v=\sqrt{v_F^2-v_1^2}\leq v_F$.
The total transfer matrix $\mathcal{T}_{C}^N$ relating the input/output voltages and currents of this meta-material transmission line is derived in Appendix \ref{app:Coulomb-Drag} and is given by
\begin{equation}
\label{eq:TCN-mat}
\mathcal{T}_C^N=\mathcal{B}_{\text{in}}\mathcal{A}^{N-2}\mathcal{B}_{\text{out}},
\end{equation}
where the matrices $\mathcal{A}$, $\mathcal{B}_{\text{in}}$ and $\mathcal{B}_{\text{out}}$ are defined in Eqs. (\ref{eq:matrix-A}), (\ref{eq:matrix-B}) and (\ref{eq:matrix-B-out}), respectively.
This result can then be used to compute the $S$ parameters.

To better understand the scattering properties of this device, it is instructive to analyze in more detail the effect of the Coulomb drag.
In particular, note that the interactions between adjacent droplets can be modeled by the transfer matrix
\begin{equation}
\label{eq:transfer-int-coulomb-drag}
\mathcal{T}_{\text{int}}=
\left(
\begin{array}{cc}
 1 & Z_{I} \\
 0 & 1
\end{array}
\right)
\left(
\begin{array}{cc}
 1 & 0 \\
 Z_G^{-1} & 1
\end{array}
\right)
\left(
\begin{array}{cc}
 1 & Z_{I} \\
 0 & 1
\end{array}
\right),
\end{equation}
with
\begin{subequations}
\label{eq:impedance-coulomb-drag}
\begin{flalign}
Z_I(\omega)&=-\frac{i}{2 \sigma_{xy}}\frac{v_F-v_1}{v} \cot \left(\frac{\omega \tau_I}{2}\right),\\
Z_G(\omega)&=-\frac{i}{2 \sigma_{xy}}\frac{v_1}{v} \cot \left(\frac{\omega \tau_I}{2}\right),
\end{flalign}
\end{subequations}
and $\tau_I=a/v$.
This transfer matrix corresponds to the equivalent circuit model shown in Fig. \ref{fig:circuit_model_coulomb_drag}.

The impedances $Z_{I,G}(\omega)$ are those of  conventional TLs terminated by an open circuit, but compared to the one in Eq.    (\ref{eq:terminal-impedance-frequency}), they have a different characteristic impedance, which depends on the intra- and inter-edge Coulomb interactions.
These impedances lead to qualitative differences compared to the case examined before, and, in particular, the impedance $Z_G(\omega)$ provides a frequency dependent connection to ground at each unit cell.
This connection generally degrades the performance of the device, however if the coupling between adjacent droplets is strong,  i.e. $v_1\rightarrow v_F$ (and consequently $v\rightarrow 0$), $Z_G(\omega)$ acts as an open circuit and $Z_I(\omega)$ as a short. 
Then, the meta-material behaves exactly as a single droplet described in Sec. \ref{sec:grounded} with a total propagation length $L_{\text{tot}}=N L_p$ ($L_p=v_0 \tau_0$ is the length of the screening electrodes of a single droplet).

\begin{figure}
\includegraphics[width=0.46\textwidth]{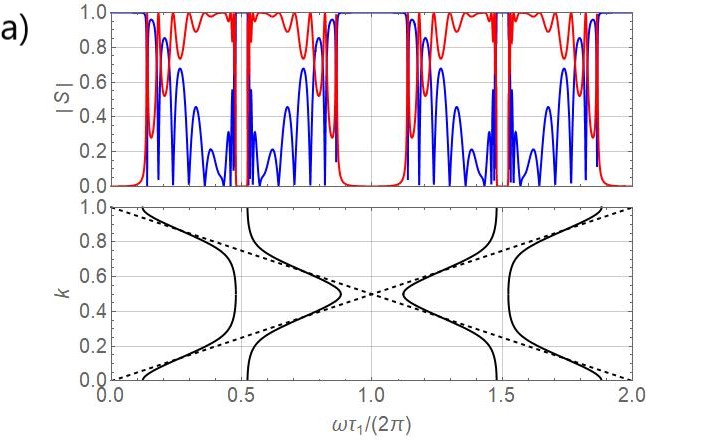}
\includegraphics[width=0.46\textwidth]{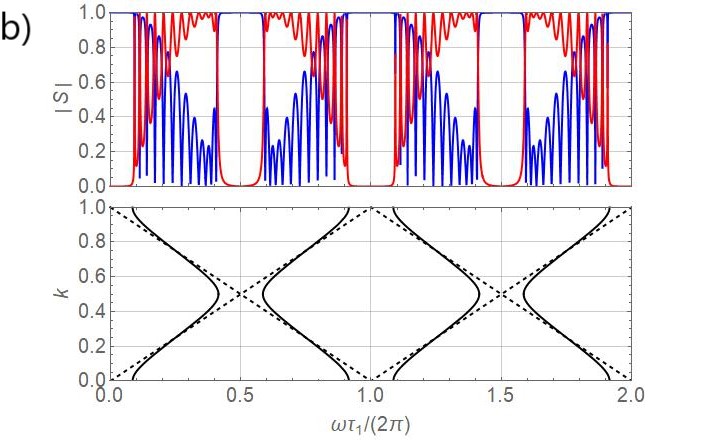}\\
\caption{Absolute value of the $S$ parameters and dispersion relation for the meta-material transmission line shown in Fig. \ref{fig:coupling_schemes_droplets} b).
The single droplets are assumed to be symmetric and matched with the external microwave circuit, i.e. $\tau_I=\tau_1$ and $\alpha=2\sigma_{xy} Z_0=1$.
The chain is composed of $N=10$ unit cells and the Coulomb coupling between adjacent droplets is parametrized by $v/v_F=0.15$.
The two plots are given for different values of $p\equiv 2\tau_0/\tau_1$, in particular in a) we used $p=1$ and in b) $p=2$.
In the top sub-plots, the blue (red) lines are $\left|S_{11}\right|$ $(\left|S_{12}\right|)$. 
In the bottom sub-plots, the band dispersions (Eq. (\ref{eq:band_coulomb_drag})) obtained with $v/v_F=0.15$ are plotted in solid lines; the dashed line is the linear dispersion obtained in the strong coupling limit $v_1\rightarrow v_F$.}
\label{fig:scattering_band_chain_cd}
\end{figure}

For finite values of $v$, new features of the scattering parameters begin to develop and, in particular, at the frequencies $\pi n/\tau_I$ and $\pi m/\tau_0$, one gets perfect reflection and suppressed transmission, as shown in Fig. \ref{fig:scattering_band_chain_cd}.

This behavior can again be understood in terms of the band structure of the chain. 
Following the same procedure discussed above, one finds the  dispersion relation 
\begin{equation}
\label{eq:band_coulomb_drag}
v_1 \cos(2\pi k)=v \sin (\omega \tau_0) \cot (\omega \tau_I)+v_F\cos (\omega \tau_0).
\end{equation}
Formally, this equation reduces to Eq.    (\ref{eq:band-structure}) when $v_1=v_F=v$, but since $v=\sqrt{v_F^2-v_1^2}$, the results above cannot be achieved for finite values of $v_1$.

In the strong coupling limit, the droplet chain has the linear dispersion relation $\omega=2\pi/\tau_0 k$ (dotted dispersion in Fig. \ref{fig:scattering_band_chain_cd}), and since $k$ is defined modulo $1$, in the first Brillouin zone there are bands touching at frequencies $\pi m/\tau_0$.
For finite values of $v$, band gaps open at these frequencies and at the frequencies $\pi n/\tau_I$, where the cotangent in Eq.    (\ref{eq:band_coulomb_drag}) diverges.
When the frequency is in the gaps, the transmission drops quickly to zero.
As $v$ approaches $v_F$, i.e. $v_1\rightarrow 0$, the bands become flatter, eventually leading to very narrow resonances.
Note that in this meta-material, when $\tau_I=\tau_1$, there are bandgaps at the central resonance frequencies $\omega_{n}$, where in the previous configurations low reflection could be achieved.
For this reason, despite the quite interesting band dispersion, a chain of Coulomb coupled droplets does not make a good TL for large $N$.

%(??Add discussion for limit of short coupling region??)
%
%(??Add discussion for asymmetric configurations??)
%
%(??Add discussion for Corbino disk??)
%
%(??Add connection to the Wilhelm-Mauch??)
%
%

\section{Conclusions and outlook}
In this paper, we discuss the possibility of manufacture low-loss transmission lines by exploiting the states localized at the edge of quantum Hall materials. 
The peculiar physics of these states offers several advantages, in particular here we focus on  their  high voltage to current ratio which guarantees a characteristic impedance of the order of the quantum of resistance.
A transmission line with an impedance of this order of magnitude offers an alternative way of achieving strong coupling between photon and spin qubits.

We analyze two possible implementations of these devices: a single QH droplet capacitively coupled to external electrodes and a meta-material transmission line with several, cascaded QH droplets.
We compute the scattering parameters of these devices to analyze different possible ways of grounding the system and to examine the effect of parasitic capacitances.
To gain additional insights, we find a simple equivalent circuit model mimicking  their response, and we observe that QH devices can also have interesting self-impedance matching properties. 
We also discuss a possible generalization of our model to account for additional plasmonic modes and Coulomb drag.

To study the meta-material transmission line, we find its effective band structure and we relate it to the power transmitted through the system.

%% Can inverted band structure be constructed - like a topological insulator?
A detailed analysis of dissipation in these devices and a quantitative analysis of the coupling to semiconductor qubits are issues that are not addressed here. More insights into these aspects can be found in \cite{HITLpt2}.

\section{Acknowledgements}
The authors thank  A. C. Mahoney,
A. C. Doherty and F. Hassler for useful discussions.
This work was supported by the Alexander von Humboldt foundation.

\begin{appendix}

\section{Coulomb Drag\label{app:Coulomb-Drag}}

Here, we explain in more detail how to derive the circuit model described in Sec. \ref{sec:meta-material-tl} for the Coulomb coupled droplet chain.

When the length of the coupling region is large compared to the distance between the two droplets, the equation of motion for the charge density in the $i$th droplet reduces to 
\begin{multline}
\label{eq:motion-emp-local-velocity-Coulomb_Drag}
i\omega\rho_i(y,\omega)= \partial_y\left(  v(y) \rho_i(y,\omega)+\right.\\
\left.v_{R}(y) \rho_{i+1}(y,\omega)+ v_{L}(y) \rho_{i-1}(y,\omega)\right).
\end{multline}
Here, we introduced the functions $v_{L,R}$ with units velocity that quantify the coupling  between adjacent droplets.

If we use a piecewise approximation for the velocity and assume a symmetric configuration, we can divide each droplet into two propagation regions $T,B$ of  length $L_p$ at the top and bottom of the droplet, where the excitation moves at a constant velocity $v_0$ and two coupling regions $R,L$ of length $a$ at the right and left of the droplet, characterized by intra- and inter-edge velocities, $v_F$ and $v_1$ respectively.
Using the results from \cite{HITLpt2} and rearranging the coordinate system to have the same (clockwise) direction in each droplet and the same origin (fixed conventionally at the boundary between region $T$ and $L$), it is quite simple to find that the general solutions for the differential equations in the four regions of the $i$th droplet are
\begin{subequations}
\label{eq:general-ansatz-CD}
\begin{align}
\label{eq:exp-ansatz}
    \rho^{T,B}_i (y,\omega )&=c^{T,B}_i e^{i \frac{\omega}{v_0} y},\\
    \rho^{R}_i (y,\omega )&=c^{+}_{i+1} e^{i \frac{\omega}{v} y}- r c^{-}_{i+1} e^{-i \frac{\omega}{v} y},\\
    \rho^{L}_i (y,\omega )&=-r c^{+}_{i} e^{-i \frac{\omega}{v} (y-L_p-2a)}+c^{-}_{i} e^{i \frac{\omega}{v} (y-L_p-2a)},
\end{align}
\end{subequations}
with $v=\sqrt{v_F^2-v_1^2}\leq v_F$ and $r=\sqrt{\frac{v_F-v}{v_F+v}}$.

The coefficients are determined by the matching conditions 
\begin{subequations}
\label{eq:matching-conditions}
\begin{align}
    v_0 \rho^{T}_i (2L_p+2a)		&=v_F \rho^{R}_i (0)+v_1\rho^{L}_{i+1} (L_p+2a),\\
    v_0 \rho^{B}_i (a)			&=v_F \rho^{R}_i (a)+v_1\rho^{L}_{i+1} (L_p+a),\\
    v_0 \rho^{T}_{i+1} (L_p+2a)	&=v_1 \rho^{R}_i (0)+v_F\rho^{L}_{i+1} (L_p+2a),\\
    v_0 \rho^{B}_{i+1} (L_p+a)	&=v_1 \rho^{R}_i (a)+v_F\rho^{L}_{i+1} (L_p+a),
\end{align}
\end{subequations}
obtained by integrating the equation of motion (\ref{eq:motion-emp-local-velocity-Coulomb_Drag}) in an infinitesimally small region enclosing the boundary of different regions, and considering the periodicity condition $\rho_T(0)=\rho_T(2L_p+2a)$; we also impose the continuity of $\rho$. 

Moreover, one can define the average currents $I^{L,R}_i$ from Eq.    (\ref{eq:current-ith-electrode}) by integrating the charge density $\rho_i$ over the regions $L,R$.
Since the resulting currents are linearly related respectively to $c_i^{\pm}$ and to $c_{i+1}^{\pm}$, using Eqs. (\ref{eq:general-ansatz-CD}) and (\ref{eq:matching-conditions}) one can straightforwardly derive for the $i$th droplet the relation
\begin{equation}
\left(\begin{array}{c}
 I_{i-1}^R \\
 I_{i}^L \\
\end{array}\right)=
\mathcal{A}
\left(\begin{array}{c}
 I_{i}^R \\
 I_{i+1}^L \\
\end{array}\right),
\end{equation}
with
\begin{subequations}
\label{eq:matrix-A}
\begin{align}
\mathcal{A}_{11}&=  \frac{2 v_F}{v_1} \left(\cos (\omega \tau_0)+\frac{v}{v_F} \cot (\omega \tau_{I}) \sin (\omega \tau_0)\right)-\mathcal{A}_{22}\\
  \mathcal{A}_{12}&=-\mathcal{A}_{21}=\frac{v_F}{v}\sin (\omega \tau_0) \cot \left(\frac{\omega \tau_I}{2}\right)+\cos (\omega \tau_0)\\
  \mathcal{A}_{22}&=-\frac{v_1}{v}\sin (\omega \tau_0) \cot \left(\frac{\omega \tau_I}{2}\right)
\end{align}
\end{subequations}
and $\tau_0=L_p/v_0$, $\tau_I=a/v$.

Note also that if we consider a sequence of $N$ equal droplets, where the first and last ones are coupled to the external electrodes, we can easily obtain  the total transfer matrix $\mathcal{T}_C^N$ in Eq. (\ref{eq:TCN-mat}).
%\begin{equation}
%\mathcal{T}_C^N=\mathcal{B}_{\text{in}}\mathcal{A}^{N-2}\mathcal{B}_{\text{out}}.
%\end{equation}
The matrices $\mathcal{B}_{\text{in/out}}$ are defined by
\begin{flalign}
&&
\left(\begin{array}{c}
 V_{\text{in}} \\
 I_{\text{in}} \\
\end{array}\right)=
\mathcal{B}_{\text{in}}
\left(\begin{array}{c}
 I_{1}^R \\
 I_{2}^L \\
\end{array}\right),
    &&
\left(\begin{array}{c}
 I_{N-1}^R \\
 I_{N}^L \\
\end{array}\right)
=
\mathcal{B}_{\text{out}}
  \left(\begin{array}{c}
 V_{\text{out}} \\
 I_{\text{out}} \\
\end{array}\right),
    &&
\end{flalign}
and can be straightforwardly obtained by considering that the voltage drive $V_{\text{in/out}}$ are applied respectively to the regions $L$ and $R$ of the initial and final droplets, and modifying accordingly the general solutions and the matching conditions.
This leads to 
\begin{subequations}
\label{eq:matrix-B}
\begin{flalign}
\begin{split}
(\mathcal{B}_{\text{in}})_{11}=&\frac{i v_F}{2 v \sigma_{xy}}\left(\cos \left(\omega \tau _0  \right) \left(\frac{v}{v_F} \cot \left(\frac{\omega\tau _1 }{2}\right) +\cot \left(\frac{\omega  \tau _I}{2}\right)\right)+ \right. \\
& \left. \sin \left(\omega \tau _0\right) \left(\cot \left(\frac{\omega \tau _1 }{2}\right) \cot \left(\frac{\omega  \tau_I}{2}\right)-\frac{v}{v_F}\right)\right)
\end{split} \\
(\mathcal{B}_{\text{in}})_{12}=&\frac{i v_1}{2 v \sigma_{xy}} \cot \left(\frac{\omega  \tau_I}{2}\right) \left(\sin (\omega \tau_0) \cot \left(\frac{\omega \tau_1}{2}\right)+\cos (\omega \tau_0)\right)\\
 (\mathcal{B}_{\text{in}})_{21}=&\mathcal{A}_{21}\\
    (\mathcal{B}_{\text{in}})_{22}=&\mathcal{A}_{22}
\end{flalign}
\end{subequations}
and to
\begin{equation}
\label{eq:matrix-B-out}
\mathcal{B}_{\text{out}}=\left(
\begin{array}{cc}
 (\mathcal{B}_{\text{in}})_{12} & (\mathcal{B}_{\text{in}})_{22} \\
 -\mathcal{A}_{22} & \mathcal{A}_{12} \\
\end{array}
\right)^{-1},
\end{equation}
with $\tau_1=L_{d}/v_d$ and $L_d,v_d$ being the length of the region coupled to the external electrodes (assumed to be the same for the input and output port) and the corresponding velocity. 

It is possible now to check that the resulting total transfer matrix $\mathcal{T}_C^N$ can be decomposed as described in Sec. \ref{sec:meta-material-tl}.

\end{appendix}
\bibliography{lit}

\end{document}